\documentclass[11pt,a4paper]{article}
\pdfoutput=1 
\usepackage{jheppub}
\usepackage[utf8]{inputenc}

\usepackage{graphicx,ulem}
\usepackage{caption}
\usepackage{subcaption}
\usepackage{hyperref}
\usepackage[T1]{fontenc}

\usepackage[table]{xcolor}
\usepackage{amssymb}
\usepackage{pifont}
\newcommand{\xmark}{\ding{55}}%

\newcommand{\be}{\begin{equation}}
\newcommand{\ee}{\end{equation}}
\newcommand{\bea}{\begin{eqnarray}}
\newcommand{\eea}{\end{eqnarray}}

\newcommand{\nn}{\nonumber\\}
\def\CC{\mathcal{C}}

\def\CN{\mathcal{N}}

\def\q{{\bf q}}

\def\im{\text{Im}}
\def\re{\text{Re}}

\title{Aspects of univalence in holographic axion models}

\author[a]{Matteo Baggioli,}
\author[b,c]{Sebastian Grieninger,}
\author[d,e]{Sa\v{s}o Grozdanov,}
\author[f]{and Zhenkang Lu}
\affiliation[a]{Wilczek Quantum Center, School of Physics and Astronomy, Shanghai Jiao Tong University, Shanghai 200240, China \& Shanghai Research Center for Quantum Sciences, Shanghai 201315.}
\affiliation[b]{Instituto de F\'isica Te\'orica UAM/CSIC, Calle Nicol\'as Cabrera 13-15, 28049 Madrid, Spain}
\affiliation[c]{Departamento de F\'isica Te\'orica, Universidad Aut{\'o}noma de Madrid, Campus de Cantoblanco, 28049 Madrid, Spain}
\affiliation[d]{Higgs Centre for Theoretical Physics, University of Edinburgh, Edinburgh, EH8 9YL, Scotland}
\affiliation[e]{Faculty of Mathematics and Physics, University of Ljubljana, Jadranska ulica 19, SI-1000 Ljubljana, Slovenia}
\affiliation[f]{Department of Physics and Astronomy, Uppsala University, Box 516, 75120 Uppsala, Sweden}

\emailAdd{b.matteo@sjtu.edu.cn}
\emailAdd{sebastian.grieninger@gmail.com}
\emailAdd{saso.grozdanov@ed.ac.uk}
\emailAdd{zhenkang.lu.5137@student.uu.se}

\preprint{IFT-UAM/CSIC-22-53} 

\abstract{Univalent functions are complex, analytic (holomorphic) and injective functions that have been widely discussed in complex analysis. It was recently proposed that the stringent constraints that univalence imposes on the growth of functions combined with sufficient analyticity conditions could be used to derive rigorous lower and upper bounds on hydrodynamic dispersion relation, i.e., on all terms appearing in their convergent series representations. The results are exact bounds on physical quantities such as the diffusivity and the speed of sound. The purpose of this paper is to further explore these ideas, investigate them in concrete holographic examples, and work towards a better intuitive understanding of the role of univalence in physics. More concretely, we study diffusive and sound modes in a family of holographic axion models and offer a set of observations, arguments and tests that support the applicability of univalence methods for bounding physical observables described in terms of effective field theories. Our work provides insight into expected `typical' regions of univalence, comparisons between the tightness of bounds and the corresponding exact values of certain quantities characterising transport, tests of relations between diffusion and bounds that involve chaotic pole-skipping, as well as tests of a condition that implies the conformal bound on the speed of sound and a complementary condition that implies its violation.    
}

\begin{document} 
\maketitle
\flushbottom

\section{Introduction}

An effective field theory (EFT) describes the dynamics of a microscopic system in a certain regime of interest by considering a subset of its original degrees of freedom (see e.g.~Refs.~\cite{Penco:2020kvy,Manohar:2018aog}). Often, this regime is the low-energy limit of a quantum field theory (QFT). Hydrodynamics is an effective field theory \cite{Kovtun:2012rj,Dubovsky:2011sj,Grozdanov:2013dba,Crossley:2015evo,Haehl:2015uoc,Jensen:2017kzi,Glorioso:2018wxw} of long-range and late-time dynamics of certain microscopic systems, which, according to its most general contemporary meaning\footnote{In this sense, hydrodynamics is not synonymous with fluid dynamics discussed, for example, in Ref.~\cite{landau}.}, applies not only to liquids and gases but can also be extended to other physical states, including solids \cite{PhysRevA.6.2401,PhysRevB.13.500,PhysRevB.13.866,Baggioli:2022pyb,PhysRev.188.898,Lucas:2017idv,PhysRevE.105.024602}. The theory of hydrodynamics is constrained by the symmetries of its underlying QFT (as every EFT). Moreover, hydrodynamics is an EFT of the globally conserved quantities, such as energy, momentum and charges (also those arising from higher-form symmetries \cite{Grozdanov:2016tdf,Grozdanov:2018ewh,Glorioso:2018kcp,Delacretaz:2019brr,Armas:2019sbe,Armas:2020bmo} or more exotic symmetries \cite{Gromov:2020yoc,PhysRevE.103.032115}), which govern the low-energy limit. More precisely, it is formulated in terms of a perturbative gradient expansion with an infinite number of unknown coefficients parametrising the lost (integrated-out) microscopic information. In equilibrium (at zeroth order in gradients), these coefficients correspond to the thermodynamic variables, such as the energy density, pressure, temperature, etc., and the relation among them is the equation of state. At higher orders in the gradient expansion, the coefficients are called the hydrodynamic transport coefficients: for example, shear and bulk viscosity, conductivity, etc. The equation of state and transport coefficients must be directly determined in terms of the microscopic dynamics and cannot be derived within the EFT description.

Since hydrodynamics describes the time-dependent collective properties of quantum systems near equilibrium, it is very natural to ask questions of the following type: How fast or slow can a quantum system relax? Similarly, how fast or slow can excitations propagate in a system? In the language of hydrodynamics, such and similar questions can be translated into the discussion of the existence, meaning, reasons for, and universality of upper and lower bounds on relaxation times, speeds, diffusion constants, conductivities, viscosities, and various other transport coefficients. For a very inexhaustive list of references on these and similar questions, see Refs.~\cite{sachdev_2011,ioffe1960non,mott1972conduction,Kovtun:2004de,Hartnoll:2014lpa,Blake:2016wvh,Zaanen:2018edk,trachenko2020minimal,Kovtun:2011np,Chafin:2012eq,Kovtun:2014nsa,Martinez:2017jjf,Lucas:2016yfl,Hartman:2017hhp,Baggioli:2020ljz,Abbasi:2020xli,Cherman:2009tw,Hohler:2009tv,Grozdanov:2015qia,Grozdanov:2015djs,Maldacena:2015waa,Kukuljan:2017xag,PhysRevLett.119.141601,PhysRevX.11.031024,PhysRevB.103.014311,doi:10.1126/sciadv.abc8662,Abbasi:2021rlp}. Importantly, many of these bounds are based on intuitive, phenomenological or heuristic arguments, which are difficult to derive from fundamental physical principles. Moreover, the aim of these bounds is often to provide universal statements about {\it all} states in `nature'. A different approach to an analogous family of problems is to temporarily abandon physical principles (at least those that we understand well) and instead concentrate on the mathematical structures that describe the physical observables of interest and explore how much information can be extracted from them.

In this spirit, Ref.~\cite{Grozdanov:2020koi} recently proposed a complex analytic method for deriving exact upper and lower bounds on properties of transport by utilising known theorems and inequalities from the theory of univalent functions, where a univalent function is a complex analytic (holomorphic) and {\it injective} function. These inequalities can be used to bound diffusion constants, the speed of sound, and higher-order coefficients that appear in the dispersion relations of hydrodynamic modes $\omega(q)$, where $\omega$ and $q$ are the frequency and the wavevector. Unlike the usual discussions of transport bounds, which focus on general arguments, these bounds should be seen as applying exact and rigorous inequalities (or bounds) to certain families of QFTs that satisfy the specified sufficient conditions. These conditions turn out to be certain univalence properties of $\omega(q)$.

What is known is that in classical (infinite-order) hydrodynamics, the dispersion relations are complex analytic (holomorphic) functions with a finite radius of convergence \cite{Grozdanov:2019kge,Grozdanov:2019uhi}\footnote{For other works on related questions, see Refs.~\cite{Bu:2014sia,Withers:2018srf,Heller:2020hnq,Heller:2020uuy,Grozdanov:2021gzh,Heller:2021oxl}.}. The univalence properties of physical theories, however, are very poorly understood. The reason for this is that we have very limited intuition into the physical meaning of injectivity of an observable. In an attempt to better understand the role of univalence in physics, in particular in EFTs such as hydrodynamics, this work is intended at contributing a few new observations to the emerging picture. In particular, we will ask questions of the following type: How large are the `typical' regions of univalence of dispersion relations $\omega(q)$ in classical hydrodynamics? Are dispersion relations also univalent beyond the disk of convergence? Moreover, how strong can the bounds that follow from univalence be? One aspect of this work is therefore to find further evidence for qualitative properties of diffusion and sound dispersion relations found in Ref.~\cite{Grozdanov:2020koi} and to establish whether the univalence methods are generically applicable for constraining transport. Another aspect is to provide new connections between univalence and the complex spectral curves introduced in Refs.~\cite{Grozdanov:2019kge,Grozdanov:2019uhi}. For these purposes, in this work, we will explicitly examine univalence in a well-known family of holographic theories with broken translational invariance called the `holographic axion models' (for a review, see Ref.~\cite{Baggioli:2021xuv}).

This paper is organised as follows: In Section~\ref{sec:Univalence}, we review Ref.~\cite{Grozdanov:2020koi} by introducing the relevant aspects of the theory of univalent functions and their applicability to establishing bounds on dispersion relations in linearised classical hydrodynamics. In Section~\ref{model}, we briefly introduce two families of holographic axion models: one with explicitly and one with spontaneously broken translational symmetry in the dual boundary QFT. In Section~\ref{sec:Diff}, we analyse the properties of energy diffusion in the longitudinal channel of the `linear axion model' with explicitly broken translations. Then, in Section~\ref{sec:solids}, we examine univalence in a theory (a `holographic solid') with spontaneously broken translations, which exhibits diffusive and sound modes in the longitudinal channel and independent sound modes in the transverse channel. Particular attention in this model is also paid to the speeds of sound of these modes in relation to the conformal speed of sound bound proposed by \cite{Cherman:2009tw,Hohler:2009tv}. Finally, in Section~\ref{sec:Summary}, we collect and summarise our main observations. Three appendices, one discussing in greater detail the local breaking of univalence in sound modes (Appendix~\ref{appendix:Cusps}), one discussing a collision between two hydrodynamic modes (Appendix~\ref{appendix:Coll}) and one showing the behaviour of the radius of convergence in an example of a holographic solid model (Appendix~\ref{appendix:R}) are added at the end of the work. 


\section{Univalent functions and the role of univalence in hydrodynamics}\label{sec:Univalence}
\subsection{Univalent functions and inequalities}\label{sec:Univalence_univalence}

Consider a function $f(z)$ defined in the complex $z$ plane or, more generally, a Riemann surface.  This function is said to be univalent in some region $z \in U$ if it is both holomorphic (analytic) and injective. The condition of injectivity means that for all pairs of $z = z_1$ and $z = z_2$ such that $z_1 \neq z_2$, $f(z_1) \neq f(z_2)$. The mathematical theory of univalent functions, which are frequently studied in the context of conformal maps, has been thoroughly developed (see for example Refs.~\cite{Ahlfors1973ConformalIT,duren2010univalent,lehto2011univalent}). A particularly powerful aspect of univalent functions is that they obey stringent constraints on their growth. Here, we briefly review some of the results (inequalities) and corresponding theorems that were used in Ref.~\cite{Grozdanov:2020koi} to constrain the hydrodynamic series. 

Showing that a holomorphic function is univalent can be a difficult task. Technically, it is very hard to compare all possible pairs of $f(z_1)$ and $f(z_2)$ and find regions where $f(z_1) \neq f(z_2)$. For this reason, numerous conditions for univalence have been developed in the literature. Commonly used conditions first derived by Nehari in 1949 \cite{Nehari1949} can be stated in terms of the Schwarzian derivative of $f(z)$ defined as\footnote{Here, and in the rest of the manuscript, primes stand for derivatives, e.g. $f'(z)\equiv df(z)/dz$.}
\begin{equation}
    \left\{f(z),z\right\} \equiv  \frac{f'''(z)}{f'(z)} - \frac{3}{2} \left( \frac{f''(z)}{f'(z)} \right)^2 .
\end{equation}
The following inequality is then a {\it necessary condition} for a holomorphic function $f(z)$ to be univalent on an open unit disk $\mathbb{D} \equiv \left\{ z\, | \,|z |< 1  \right\}$:
\begin{equation}
    \left| \left\{f(z),z\right\}\right| \leq \frac{6}{\left(1 - |z|^2 \right)^2},
\end{equation}
whereas the inequality 
\begin{equation}
   \left| \left\{f(z),z\right\} \right| \leq \frac{2}{\left(1 - |z|^2 \right)^2}
\end{equation}
is a {\it sufficient condition} for univalence on $\mathbb{D}$. For adaptations of the latter statement, see for example Ref.~\cite{aharonov2013sufficient}.

A different sufficient condition that can be more easily adapted to the language of hydrodynamics was used in Ref.~\cite{Grozdanov:2020koi}. The condition was originally derived by Noshiro in 1934 \cite{noshiro1934theory} and by Warschawski in 1935 \cite{warschawski1935higher}, and can be stated as follows. If a holomorphic function $f(z)$ has the property that
\begin{equation}\label{Noshiro}
    \re f'(z) > 0 , ~\text{for}~z \in U ,
\end{equation}
where $U$ is an open convex region in the complex plane, then $f(z)$ is univalent in $U$. It is easy to understand how this statement relates to univalence. Consider two points $z_1$ and $z_2$ in a convex region $U$. Then the line segment connecting the two points $\gamma(\tau) = z_1 + (z_2 - z_1) \tau$, where $0\leq\tau\leq 1$ also lies entirely in $U$ (by assumption of convexity). Hence, it follows that for $z_1 \neq z_2$,
\begin{equation}
    f(z_1) - f(z_2) = \left(z_1 - z_2 \right)\left[\int_0^1 \re f'(\gamma(\tau)) \,d\tau + i \int_0^1 \im f'(\gamma(\tau)) \,d\tau    \right] \neq 0  
\end{equation}
if $\re f'(z) > 0 $. As can be seen, one could also equivalently use the imaginary part in the univalence condition. Assuming now that univalence has been established in $U$, then the region can be mapped to an open unit disk $\mathbb{D}$ in the complex $\zeta$-plane via a conformal map $\varphi: U \to \mathbb{D}$ so that $\zeta = \varphi(z)$ and $z = \varphi^{-1} (\zeta)$. This is ensured by the {\it Riemann mapping theorem}. The function $f(\zeta)$ is now univalent in $\zeta \in \mathbb{D}$. We further normalise $f(\zeta)$ so that $f(0) = 0$ and $f'(0) = 1$. Univalence then implies a number of powerful inequalities on the function. Firstly, it implies the {\it growth theorem}:
\begin{equation}\label{GrowthTheorem}
    \frac{|\zeta|}{\left(1 + |\zeta|\right)^2} \leq |f(\zeta)| \leq \frac{|\zeta|}{\left(1 - |\zeta|\right)^2},
\end{equation}
as well as a number of related statements regarding the growth of the derivatives of $f$. Another, even more striking statement that stems from univalence is the so-called {\it de Branges theorem}, originally known as the {\it Bieberbach conjecture} (for the 1985 proof of the statement, see Ref.~\cite{branges1985}). It can be stated as follows. A univalent function $f(\zeta)$ in $\mathbb{D}$ with the normalisation $f(0) = 0$ and $f'(0) = 1$ admits a convergent power series expansion:
\begin{equation}\label{fSeries}
    f(\zeta) = \zeta + \sum_{n=2}^\infty b_n \zeta^n.
\end{equation}
The de Branges theorem then constrains each of the individual coefficients of the series as
\begin{equation}\label{deBranges}
|b_n| \leq n, ~ \text{for all}~ n \geq 2.
\end{equation}

An even more restrictive set of inequalities can be established if the univalent function on the open unit disk satisfies a further condition
\begin{equation}\label{MacGregor}
\re f'(\zeta) > 0, ~ \text{for}~ \zeta \in \mathbb{D}.
\end{equation}
It is important to note that Eq.~\eqref{Noshiro}, which applies to general convex domains, does {\it not} imply Eq.~\eqref{MacGregor}. However, for an important class of special cases whereby $\re f'(z) > 0$ on the entire convergence disk $|z| < R$, the property $\re f'(\zeta) > 0$ is automatically implied. This is because the relevant conformal map from the original univalent domain to a unit disk $|\zeta| < 1$ is a special case of the M\"{o}bius map, which is a simple rescaling of $z$. For functions that do satisfy Eq.~\eqref{MacGregor}, MacGregor showed in 1962 that \cite{macgregor1962functions}
\begin{equation}\label{GrowthMacGregor}
        -|\zeta| + 2\ln \left(1 + |\zeta| \right) \leq |f(\zeta)| \leq  -|\zeta| - 2\ln \left(1 - |\zeta| \right), 
\end{equation}
and that the coefficients of such a series \eqref{fSeries} satisfy a stricter set of inequalities than \eqref{deBranges}:
\begin{equation}\label{MacGregorCoeffs}
    |b_n| \leq \frac{2}{n}, ~ \text{for all}~ n \geq 2.
\end{equation}
The latter two equations are a simple consequence of a lemma proven by Caratheodory (for the statement of the lemma and relevant references, see Ref.~\cite{macgregor1962functions}). We will frequently refer to the equations \eqref{GrowthMacGregor} and \eqref{MacGregorCoeffs} as the {\it log} bounds.

Ref.~\cite{Grozdanov:2020koi} showed how these inequalities can be used in the context of effective field theory, in particular, hydrodynamics. Provided that one can establish the univalence of an observable that is naturally expanded in a series, then the coefficients of the series can be bounded. In hydrodynamics, such a procedure can bound for example the coefficients of the dispersion relations. Along similar lines, one can also put bounds on the coefficients of the Wilsonian effective action which are entering into the scattering amplitudes computed in QFT~\cite{Haldar:2021rri}.

\subsection{Hydrodynamic dispersion relations}\label{sec:Univalence_hydro}

In this work, as in \cite{Grozdanov:2020koi}, we focus on hydrodynamic dispersion relations, namely, the diffusive and sound dispersion relations. Their generic form follows from the Puiseux theorem applied to the hydrodynamic constitutive relations and the resulting conservation laws \cite{Grozdanov:2019kge,Grozdanov:2019uhi} (see also \cite{Bu:2014ena,Bu:2014sia,Bu:2015bwa}). These considerations lead to a specific structure of the {\it complex spectral curve} in Fourier space $P(\omega, \q^2)$, where $\omega$ is the frequency and $\q$ is the wavevector (sometimes called momentum), which is a spatial $3$-vector. The fact that $P$ depends on $\q^2$ is therefore a result of the assumption that, here, for simplicity, we consider only rotationally invariant systems. The modes in the spectrum are the solutions of the equation $P(\omega, \q^2) = 0$, typically solved for $\omega(\q^2)$ and expressed as a power series in the (scalar) expansion parameter $\q^2$ --- the small wavevector. The series has a finite radius of convergence determined by critical points of the curve; in the simplest case with $\partial_\omega P(\omega, \q^2) = 0$ and $\partial^2_\omega P(\omega, \q^2) \neq 0$ \cite{Grozdanov:2019kge,Grozdanov:2019uhi}.\footnote{For earlier numerical work studying a specific holographic example of a charged fluid where the convergence of a hydrodynamic series was explicitly observed, as was the fact that the radius of convergence is limited by a collision with a non-hydrodynamic mode at imaginary momentum, see Ref.~\cite{Withers:2018srf}. For an explanation of certain details of \cite{Grozdanov:2019uhi} pertaining to the factorisation of spectral curves, `level-crossing' and `level-touching', see also Ref.~\cite{Grozdanov:2021gzh}. For statements regarding the convergence of linearised hydrodynamics in position space, see Ref.~\cite{Heller:2020uuy}, and for a recent discussion of non-linear flows in position space, see Ref. \cite{Heller:2021oxl}.} For the diffusive mode, we can choose $z\equiv \q^2$ and also define $q \equiv \sqrt{\q^2} = |\q|$ for future use. The dispersion relation then has the form
\begin{equation}
\omega (z \equiv {\bf q}^2) = - i \sum_{n=1}^\infty c_n z^n  , \label{WDiff}
\end{equation}
where $c_1 = D$ is the diffusivity and, in general, all $c_n \in \mathbb{R}$. While the sign of $c_1 = D \geq 0$ is fixed by the positivity of the entropy production, the signs of the remaining coefficients $c_{n\geq 2}$ have not been fixed. This means that $\omega$ is purely imaginary for real $z$ and the mode is not propagating. The series \eqref{WDiff} converges on the disk of radius $R$:
\begin{equation}
    \mathbb{D}_R = \{z\,|\,|z| < R \}.
\end{equation}
The value of $R$ is (at least)\footnote{By `at least', we refer to the question of `level-crossing' versus `level-touching' discussed in Ref.~\cite{Grozdanov:2019uhi}. Specifically, what `level-touching' implies in such a case is that the radius of convergence is determined by a sub-leading critical point where `level-crossing' occurs. See for example the analysis of the BTZ black hole in \cite{Grozdanov:2019uhi}. For further discussions of these issues, see also Refs.~\cite{Grozdanov:2021gzh} and \cite{Heller:2020uuy}.} as large as $R = |z_*|$, where $z_*$ is determined  by the leading critical point of the associated complex spectral curve \cite{Grozdanov:2019kge,Grozdanov:2019uhi}. Note that the choice of $z = q^2$ is only a matter of convenience. This will be the choice that we make in Section~\ref{sec:Diff}. However, in Section~\ref{sec:solids}, where we also study diffusion, but one that couples to sound modes, it will prove more useful to use the complexified $z \equiv q$ instead.

For sound modes, it is most convenient for present purposes to use the complex variable $z\equiv \sqrt{\q^2}$. The pair of the dispersion relations  then
\begin{equation}
\omega_\pm  (z \equiv \sqrt{\q^2}) = - i \sum_{n=1}^\infty a_n \,e^{\pm \frac{i\pi n}{2}} z^n ,  \label{WSound}
\end{equation}
where $a_1 = v_s$ is the speed of sound and also with all $a_n \in \mathbb{R}$. For real $z$, these propagating dispersion relations have alternating real and imaginary parts and $\omega(z)$ is complex. The real part determines the propagating nature of the sound wave, while the imaginary part describes its attenuation. These properties are completely fixed by the order of the critical point around which the series are expanded \cite{Grozdanov:2019kge,Grozdanov:2019uhi}. In what will follow below, when we discuss the univalence of the sound mode, we will choose $\omega(z) = \omega_+(z)$. Here, we also define the concept of the group velocity\footnote{Strictly speaking, this corresponds to a physical velocity only when $z \in \mathbb{R}$.}, which will be particularly useful for the sound mode,
\begin{equation}\label{GroupVel}
    v_g(z) \equiv \omega'(z).
\end{equation}
Again, we consider this quantity for any complexified $z\in \mathbb{C}$.

\begin{figure}[h!]
\centering
 \includegraphics[width=1\textwidth]{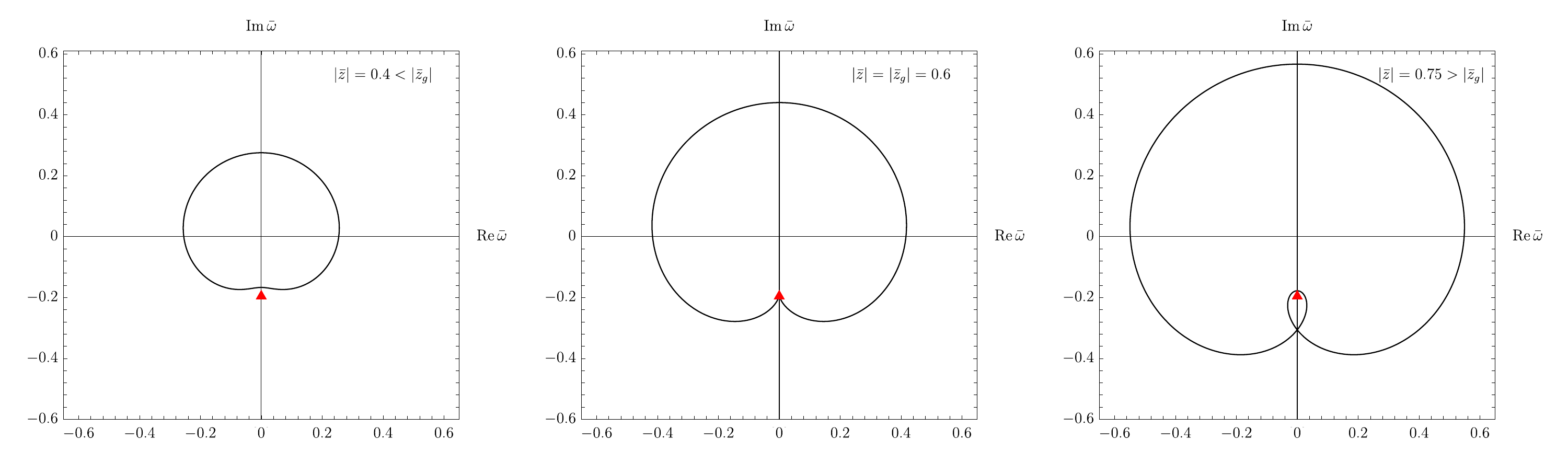}
\caption{\label{fig:CuspN4SYM}
Hydrodynamic sound dispersion relation $\bar\omega(\bar z)$, where $\bar \omega \equiv \omega / (2\pi T)$ and $\bar z \equiv z / (2\pi T)$, in the $\CN = 4$ SYM theory showing the development of a cusp at $\bar z = \bar z_g = - 0.603 i$. The curve is formed by writing $\bar z = |\bar z| e^{i \phi}$ and varying $\phi \in [0, 2 \pi]$. The value of $\bar \omega(\bar z_g)$ is depicted with a red triangle. After the formation of the cusp (i.e., for $|z| > |z_g|$) the curve has a self-intersection.}
\end{figure}

As argued in Ref.~\cite{Grozdanov:2020koi}, hydrodynamic dispersion relations have a finite range of univalence $U$ that includes the origin $z=0$. This statement follows from the fact that since $\omega'(z) \neq 0$, $\omega(z)$ are by the implicit function theorem invertible and thus bijective at the origin. One can therefore always use univalence methods to make statements about hydrodynamics. Moreover, what is clear from the definition of the group velocity \eqref{GroupVel} is that if 
\begin{equation}\label{local_cond}
v_g(z_g) = \omega'(z_g) = 0,
\end{equation} 
for some $z_g \in \mathbb{C}$, then this point certainly signals the end of univalence (see discussion in \cite{Grozdanov:2020koi}). We will refer to such points as causing local violation of univalence. Moreover, in the language of \cite{Grozdanov:2019kge,Grozdanov:2019uhi}, such points have a clear interpretation from the point of view of the spectral curve solutions $\omega(z)$. They can be understood as cusps that form in the $\omega(z)$ curve at $z_g$ when plotted in the complex $\omega$ plane. We show this for the simple case of the $\CN = 4$ supersymmetric Yang-Mills (SYM) theory with an infinite number of colours and at infinite coupling in Figure~\ref{fig:CuspN4SYM}.

For sound modes, where these points of local univalence breaking play an important role, we observe that they occur for purely imaginary values of $z_g$. A simple argument that supports this observation can be made by using the recursive reasoning of the Newton-Raphson method. We discuss the relevant details in Appendix~\ref{appendix:Cusps}.

\subsection{Bounds on hydrodynamics}\label{sec:Univalence_hydrobounds}

To review the steps used in Ref.~\cite{Grozdanov:2020koi} for deriving the hydrodynamic bounds, we focus on the case of diffusion. Sound works in an analogous manner. In order to find bounds on $D$, we need to know some region of univalence $U$ of the dispersion relation $\omega(z)$ so that a map to $\mathbb{D}$ in the $\zeta$ plane can be constructed. In addition, we also need to know the value of the dispersion relation at a point $z_0 \in U$. We denote this point as $\omega_0 \equiv \omega(z_0)$. In terms of the complex variable $\zeta$ after the conformal map, $\zeta_0 \equiv \varphi(z_0)$. It is also important to stress that for all conformal maps used here, we take $\varphi(0) = 0$. We then define a new function $f(\zeta)$ in terms of the diffusive dispersion relation $\omega(z)$ from Eq.~\eqref{WDiff}:
\begin{equation}
    f(\zeta) \equiv \frac{i \omega(\varphi^{-1}(\zeta))}{D \,\partial_\zeta \varphi^{-1}(0)},
\end{equation}
which has, by construction, a power series expansion of the form of Eq.~\eqref{fSeries}. Then, using the growth theorem \eqref{GrowthTheorem}, we find
\begin{equation}\label{D_bound_Uni}
    \frac{|\omega_0| \left(1-|\zeta_0|\right)^2}{|\zeta_0| \left|\partial_\zeta \varphi^{-1}(0) \right|} \leq D \leq \frac{|\omega_0| \left(1 + |\zeta_0|\right)^2}{|\zeta_0| \left|\partial_\zeta \varphi^{-1}(0) \right|} .   
\end{equation}
 If $\re f'(\zeta) > 0$, then we can instead use the inequalities in \eqref{GrowthMacGregor} to find 
\begin{equation}\label{D_bound_MacG}
    \frac{|\omega_0| }{ \left|\partial_\zeta \varphi^{-1}(0) \right| \ln \left[\frac{1}{\left(1 - |\zeta_0| \right)^2} e^{-|\zeta_0|}   \right] } \leq D \leq   \frac{|\omega_0| }{ \left|\partial_\zeta \varphi^{-1}(0) \right| \ln \left[\left(1 + |\zeta_0| \right)^2 e^{-|\zeta_0|}   \right] } .  
\end{equation}
Higher-order coefficients can be bounded by applying either Eq.~\eqref{deBranges} or Eq.~\eqref{MacGregorCoeffs} (depending on whether only $\re f'(z)>0$ or also $\re f'(\zeta) > 0$) to the coefficients $b_{n\geq 2}$, which can be computed directly from $c_n$ and the conformal map $\varphi(z) = \zeta$.
  
We note that our discussion below Eq.~\eqref{MacGregor} can now be phrased as follows. If for diffusion, $\re \left[i \omega'(z)\right]>0$ and $\varphi$ is a map from a disk centred at $z=0$ to a disk centred at $\zeta = 0$ (i.e. a simple rescaling M\"{o}bius transformation), then $\re f'(\zeta) > 0$ is also true. Hence, Eq.~\eqref{D_bound_MacG} can always be used in such cases instead of the standard univalence inequalities in Eq.~\eqref{D_bound_Uni}. 

Finally, we comment on the question of how we can think of the cases for which $U$ that we wish to use is larger than $\mathbb{D}_R$. The statement is that we can use the above bounds despite the fact that the hydrodynamic series fails to converge outside of $\mathbb{D}_R$. This is because the dispersion relation $\omega(z)$ for all $z\in U$ can be thought of as the analytically continued series representation of $\omega(z)$ from $z\in\mathbb{D}_R$. For concreteness, consider again the diffusive hydrodynamic series \eqref{WDiff}. For any $U$ that contains $\mathbb{D}_R$  (recall that $\omega(z)$ must be holomorphic in $U$), there exists an analytic continuation --- a conformal map $y = \psi(z)$ and $z =\psi^{-1}(y)$ with the property $\psi(0) = 0$ --- that maps the original region $U$ in the $z$ plane into a disk $\mathbb{D}$ in the $y$ plane of which the radius is still determined by the obstruction to convergence of the original $\omega(z)$ series. The following series representation
\begin{align}
    \omega(z) = \omega\left(\psi^{-1}(y)\right) = - i \sum_{n=1}^\infty \tilde c_n\, y^n \equiv \tilde \omega(y)
\end{align}
now converges for all $y \in \psi(U)$. In other words, the convergent domain of the series $\tilde \omega(y)$ covers the entire original univalent domain $z \in U$ and any coefficient $\tilde c_n$ depends only on the hydrodynamic coefficients $c_{n \leq N}$ and the map $\psi$. For example, using $c_1 = D$,
\begin{align}
 \tilde c_1 &= D \partial_y \psi^{-1}(0), \\
 \tilde c_2 &= c_2 \left(\partial_y \psi^{-1}(0)\right)^2 + \frac{1}{2} D \partial^2_y \psi^{-1}(0).
\end{align}
We can now apply the entire univalence discussion directly to $\tilde \omega(y)$. We define a conformal map $\gamma(y) = \zeta$, and $y = \gamma^{-1}(\zeta)$, from the convergence disk in the $y$ plane to a unit disk $\mathbb{D}$ in the $\zeta$ plane and the function
\begin{equation}
    f(\zeta) \equiv \frac{i \tilde\omega\left(\gamma^{-1}(\zeta) \right) }{\tilde c_1 \partial_\zeta \gamma^{-1}(0)} = \frac{i \omega\left( \psi^{-1} \left( \gamma^{-1} (\zeta) \right)\right) }{D \partial_y \psi^{-1}(0) \partial_\zeta \gamma^{-1}(0)} = \frac{i \omega\left(  \varphi^{-1} (\zeta) \right) }{D \partial_\zeta\varphi^{-1}(0)}  = \zeta + \sum_{n=1}^\infty b_n \zeta^n,
\end{equation}
where $\zeta = \varphi(z) = \gamma(\psi(z))$ with the inverse $z = \varphi^{-1}(\zeta) = \psi^{-1}(\gamma^{-1}(\zeta))$. Here, a composition of two conformal maps is a conformal map mapping a univalent function into a univalent function. Hence, what the expression shows is that the series coefficients $b_n$ in the unit disk $\zeta \in \mathbb{D}$ are the same regardless of whether they are constructed from $\omega(z)$ or $\tilde\omega(y)$. All $b_n$ can be fixed in terms of derivatives of the conformal map(s) and the hydrodynamic coefficients $c_n$. We can therefore use the bounds regardless of whether the point $\omega_0(z_0)$ lies inside or outside of $\mathbb{D}_R$, so long as it lies in $U$.

\section{Holographic axion models and their hydrodynamic spectra}\label{model}

The holographic axion models are a class of bottom-up holographic setups in which translational invariance of the dual field theory is explicitly or spontaneously broken while the homogeneity of the bulk geometry is retained. The simplest such model is the so-called \textit{linear axion model} with explicitly broken translations introduced in \cite{Andrade:2013gsa}, which was later generalised in \cite{Baggioli:2014roa,Alberte:2015isw} (for a review, see Ref.~\cite{Baggioli:2021xuv}). Here, we briefly describe certain features of these models, in particular, of the aspects relevant to this work: the hydrodynamic excitations. For an exposition of various other detailed features of the hydrodynamic excitations in these theories and their corresponding effective field theory descriptions, see Ref.~\cite{Baggioli:2022pyb}. 

The action for this class of models with four bulk (or, three boundary) spacetime dimensions is
\begin{equation}
S=\int d^4x\sqrt{-g}\left[R-2\Lambda- V(X,Z)\right],
\label{action}
\end{equation}
where $R$ is the Ricci scalar and $\Lambda = - 3/L^2$ the negative cosmological constant that depends on the AdS radius $L$. The invariants $X$ and $Z$ are expressed in terms of the matrix $\mathcal{I}^{IJ}=g^{\mu\nu}\partial_\mu\phi^I\partial_\nu\phi^J$ that depends on the scalar fields $\phi^I$ (the `axions'), where the indices $I$ and $J$ run over the spatial directions of the boundary field theory, i.e.,~$I,J = \{1,2\}$. Explicitly, $X=\frac{1}{2}\,\mathrm{Tr}\,\mathcal{I}^{IJ}$ and $Z=\mathrm{det}\,\mathcal{I}^{IJ}$. Finally, $V$ is a scalar function that has to obey certain consistency constraints described in Ref.~\cite{Baggioli:2014roa}. 

The classical background bulk solution for the axions is given by
\begin{equation}
    \phi^I =m x^I . \label{eq:bulkscalar}
\end{equation}
The action \eqref{action} is invariant under the global shift symmetry $\phi^I\rightarrow \phi^I+b^I$, with $b^I$ constant. However, the solution \eqref{eq:bulkscalar} is not invariant under translations $x^i \rightarrow x^i+a^i$, with $a^i$ constant, thereby breaking the translational symmetry. For this reason, the background solution \eqref{eq:bulkscalar} breaks the pair of spatial translations and the internal global shifts to their diagonal subgroup, which `correlates' the spatial and the internal indices. This is analogous to what happens in the (boundary) effective field theory constructions of elasticity discussed in Ref.~\cite{Nicolis:2015sra}. Whether the breaking of translations is explicit or spontaneous from the point of view of the dual field theory depends on the specific form of the potential $V$ in the action \eqref{action} and the consequent expansion of the axion fields close to the asymptotic AdS boundary.\footnote{Spontaneous symmetry breaking could also be realised by modifying the boundary conditions of the bulk axion fields \cite{Armas:2019sbe,Ammon:2020xyv}. Unfortunately, this seems to result in dual field theories with a negative shear modulus and therefore dynamical instabilities. We will not consider this option here.}. The parameter $m$ determines the `strength' of the translational symmetry breaking.\footnote{See Ref.~\cite{Alberte:2015isw} for a discussion of how this theory is related to a theory of massive gravity and, in particular, how $m$ is related to the mass of gravitons.}

In all relevant cases, the background geometry is captured by the following ansatz:
\begin{equation}
\label{backg}
ds^2=\frac{1}{u^2} \left[-f(u)dt^2-2dt du + dx^2+dy^2\right] ,
\end{equation}
where $\{t,u,x,y\}$ are the infalling Eddington-Finkelstein coordinates and the AdS length scale has been fixed to unity, $L=1$. In our notation, the AdS boundary is located at $u=0$ and the black brane horizon will be fixed to lie at $u=u_h$. The emblackening factor appearing in Eq.~\eqref{backg} takes the following form:
\begin{equation}\label{backf}
f(u)= u^3 \int_u^{u_h} dv\left[ \frac{3}{v^4} -\frac{V(m^2v^2,m^4v^4)}{2 v^4} \right].
\end{equation}
The corresponding temperature of the dual theory is then
\begin{equation}
T=-\frac{f'(u_h)}{4\pi}=\frac{6 -  V\left(m^2 u_h^2,m^4u_h^4 \right)}{8 \pi u_h}, \label{eq:temperature}
\end{equation}
and the entropy density is $s=2\pi/u_h^2$. In the rest of the manuscript, we will always set $u_h=1$.

\subsection{The linear axion model}\label{sec:LinearAxion_Intro}

The \textit{linear axion model}, which first appeared in the context of holography in Ref.~\cite{Andrade:2013gsa} (see also \cite{Taylor:2014tka} for its first direct generalization and \cite{Caldarelli:2016nni} for the discussion of the model's thermodynamic properties), refers to the linear potential choice in Eq.~\eqref{backg}: namely, $V(X)=X$. This is the simplest model that can account for momentum dissipation in the dual field theory. With the specific choice of the potential $V$, the background profile for the axion fields \eqref{eq:bulkscalar} corresponds in the standard (Dirichlet) quantisation to an external spatially dependent source for the dual scalar operators. For this reason, the boundary sources, which are directly related to $\phi^I$, are responsible for explicit breaking of translational invariance in the dual field theory and the introduction of a finite momentum relaxation rate \cite{Davison:2013jba}.

The spectrum of hydrodynamic excitations in systems without momentum relaxation includes transverse diffusive and longitudinal sound modes. Such systems are holographically dual to the pure Einstein gravity in the bulk. Since momentum is no longer a conserved quantity in the dual of the linear axion model, the transverse diffusive hydrodynamic mode acquires a finite relaxation rate and therefore becomes gapped. Its dispersion relation changes from the standard diffusive expression \eqref{WDiff} (i.e., $\omega=-i D q^2 + \ldots $, with the diffusivity $D = \eta / (\varepsilon + p )$, where $\varepsilon$, $p$ and $\eta$ are the energy density, pressure and the shear viscosity) to a Drude-like form $\omega=-i \Gamma + \ldots$, where the relaxation rate scales as $\Gamma \sim m^2$ \cite{Davison:2013jba}. The spectrum of transverse fluctuations therefore contains {\it no} hydrodynamic modes. 

On the other hand, because of energy conservation, the longitudinal channel does retain a gapless hydrodynamic excitation, which is due to the momentum relaxation of a diffusive mode instead of a familiar sound mode. We will refer to this mode as the {\it energy diffusion} (see Ref.~\cite{Davison:2014lua} for details). This is because in the \textit{incoherent limit}, where the strength of momentum dissipation $\Gamma$ is the dominant scale in the system ($q\ll T \ll \Gamma$), its dispersion relation becomes that of energy diffusion $\omega=-i D_e q^2 + \ldots$, with the diffusivity determined by the Einstein relation $D_e= \kappa / c_V$. Here, $\kappa$  and $c_V$ are the thermal conductivity and the heat capacity at constant volume, respectively. At intermediate values of momentum dissipation, $m/T \sim 1$, the model displays the so-called \textit{k-gap} behaviour \cite{Baggioli:2019jcm} in which the diffusive mode collides with the first non-hydrodynamic mode at a certain critical momentum signalling the breakdown of hydrodynamics expansion and the need for quasihydrodynamic considerations \cite{Grozdanov:2018fic}, in which the long-lived gapped modes are included in the low-energy effective theory \cite{Grozdanov:2018fic,Baggioli:2020loj} (see also \cite{Stephanov:2017ghc}). This diffusive mode exhibited by the longitudinal channel is the hydrodynamic mode of which the univalence properties will be studied in detail in Section~\ref{sec:Diff}.

\subsection{Holographic solids}\label{sec:Solids_Intro}

A second sub-class of the holographic axion models is that of the so-called \textit{holographic solids}. With this jargon, we refer to homogeneous axion models which spontaneously (instead of explicitly) break translational invariance in the dual field theory. Such theories, which preserve both energy and momentum, describe solid systems and their spectra contain massless Goldstone modes, which can be associated with phonons. Holographic solids were introduced in \cite{Alberte:2017oqx} and, in their simplest incarnation, they can be realised by a power-law potential $V(X)=X^N$ with $N>5/2$. For such sufficiently high powers of $N$, the axions in Eq.~\eqref{eq:bulkscalar} can be associated with expectation values of dual scalar operators rather than their sources (see Ref.~\cite{Alberte:2017oqx}  for details). For simplicity and for the fact that the important qualitative features of the holographic solid models are independent of the specific choice of $N$, hereon, we will set $N=3$.

In this scenario, the hydrodynamic (gapless) spectrum is more complicated than in the linear axion model. This is because of the appearance of additional massless excitations mentioned above, which are the transverse and the longitudinal Goldstone modes. In the transverse sector, the Goldstone mode couples to the conserved transverse momentum to produce a propagating sound mode known as the shear (or transverse) sound. The speed of propagation of this mode is determined by the rigidity of the system, i.e., by its elastic shear modulus \cite{Alberte:2017oqx}. The parameter $m$ now controls the `strength' of elasticity rather than the momentum relaxation rate. Interestingly, by imposing obvious constraints such as subluminality and using the conditions forced by conformal symmetry, one finds that the speed of transverse sound $v_T$ is at most the conformal speed \cite{Esposito:2017qpj,Baggioli:2020ljz}: 
\begin{equation}\label{TsoundSpeed}
0 \leq v_T \leq v_c \equiv \sqrt{\frac{1}{d-1}} ,
\end{equation} 
where $d$ is the number of the boundary spacetime dimensions. 

In the longitudinal channel, the standard sound mode persists but its speed is much larger because of the presence of finite elastic moduli \cite{Ammon:2019apj}. Moreover, a new diffusive mode known as \textit{crystal diffusion} appears in the spectrum. As shown in Ref.~\cite{Donos:2019txg}, its origin is rooted in the global shift symmetry of the model. Importantly, because of conformal invariance \cite{Esposito:2017qpj}, the speeds of the transverse sound $v_T$ and the longitudinal sound $v_L$ always satisfy the constraint 
\begin{equation}
v_L^2=v_c^2 + v_T^2.
\end{equation} 
In consequence, the speed of the longitudinal sound mode in these models is therefore always above the conformal value and below the speed of light (set to $c =1$) \cite{Baggioli:2020ljz}:
\begin{equation}\label{LsoundSpeed}
v_c \leq v_L \leq 1 .
\end{equation}
The full hydrodynamic structure of these models is well captured by relativistic viscoelastic hydrodynamics discussed in Ref.~\cite{Armas:2019sbe,Armas:2020bmo}, which has been shown to match the holographic results \cite{Ammon:2020xyv,Baggioli:2020edn}. The univalence properties of transverse and longitudinal modes will be discussed in Section~\ref{sec:solids}.

\subsection{Summary of hydrodynamic modes present in the spectra}

In the last part of this section, we summarise the types of hydrodynamic modes exhibited by the axion models described above that we consider in this work. We also compare the spectrum to that of the neutral thermal conformal field theories (CFTs) dual to pure Einstein gravity in the bulk, such as for example the $\CN = 4$ SYM theory in $d=4$. Those arise in the limit of $m\to 0$ in which the axion fields decouple from gravity and can be consistently set to zero. All hydrodynamic types of modes are collected in Table~\ref{tab1}. 

\begin{table}[h!]
\centering
\begin{tabular}{ |p{4cm}|p{4cm}|p{4cm}|  }
\hline
\multicolumn{3}{|c|}{\cellcolor{blue!08} \textbf{Spectrum of hydrodynamic modes}} \\
\hline
\cellcolor{gray!10} Model & \cellcolor{gray!10} Transverse channel & \cellcolor{gray!10} Longitudinal channel\\
\hline
\hline
pure gravity&  shear diffusion & longitudinal sound\\
\hline
linear axion model & \xmark & energy diffusion\\
\hline
holographic solid &  shear sound & longitudinal sound and\\
 &  & crystal diffusion\\
 \hline
\end{tabular}
\caption{Hydrodynamic (gapless) excitations that exhibited by the models considered in this work: the pure Einstein gravity theory with a Schwarzschild black brane solution, the linear axion model with explicitly broken translations and the holographic solid model with spontaneously broken translations. The symbol \xmark $\,$ indicates the absence of hydrodynamic modes in the particular channel.}
\label{tab1}
\end{table}

\section{The linear axion model with explicitly broken translations:\\ energy diffusion}\label{sec:Diff}

In the linear axion model introduced in Section~\ref{sec:LinearAxion_Intro}, it is easy to check by explicit calculation that as in Ref.~\cite{Grozdanov:2020koi}, the dispersion relations of energy diffusion in the longitudinal channel are univalent on the disk of convergence $|z| < R$ for all values of the ratio $m/T$, where the temperature $T$ is given by Eq.~\eqref{eq:temperature}. In in our choice of units ($u_h=1$), it becomes
\begin{equation}\label{Temperature}
T= \frac{6-m^2}{8 \pi} .
\end{equation}
Then, the dimensionless ratio, which we will use throughout, is
\begin{equation}\label{moverT}
\frac{m}{T} = \frac{8\pi m}{ 6-m^2 }. 
\end{equation} 
It is purely parametrised by $m \in [0, \sqrt{6}] $. To compute various properties of the longitudinal channel spectrum in the dual quantum field theory, we analyse the gauge-invariant bulk equation of motion, which can be found for example in Refs.~\cite{Davison:2014lua,Blake:2018leo}. In particular, we solve Eq.~(3.5) in Ref.~\cite{Blake:2018leo} by using the methods developed in Ref.~\cite{Kovtun:2005ev}.  

\begin{figure}[h!]
\centering
 \includegraphics[width=0.45\textwidth]{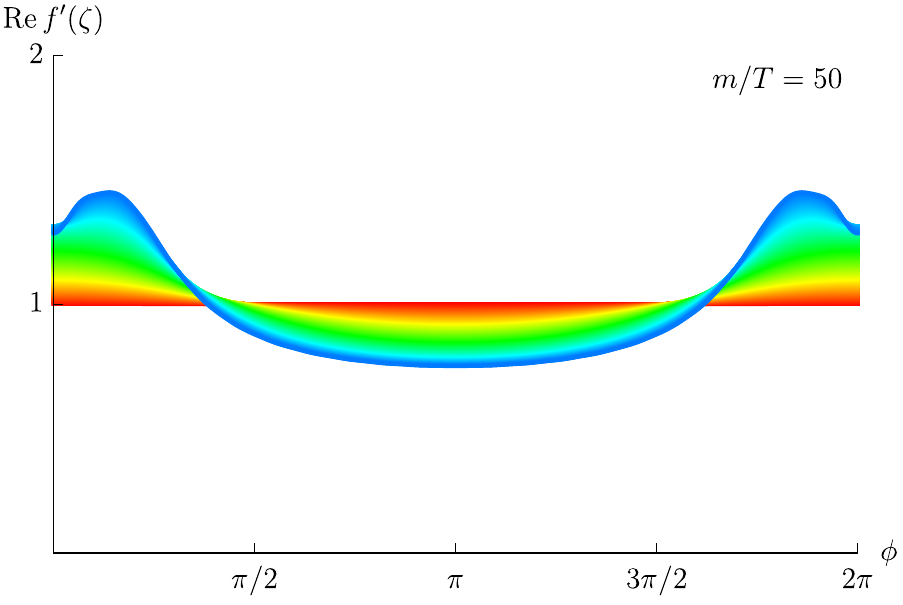} \qquad
 \includegraphics[width=0.45\textwidth]{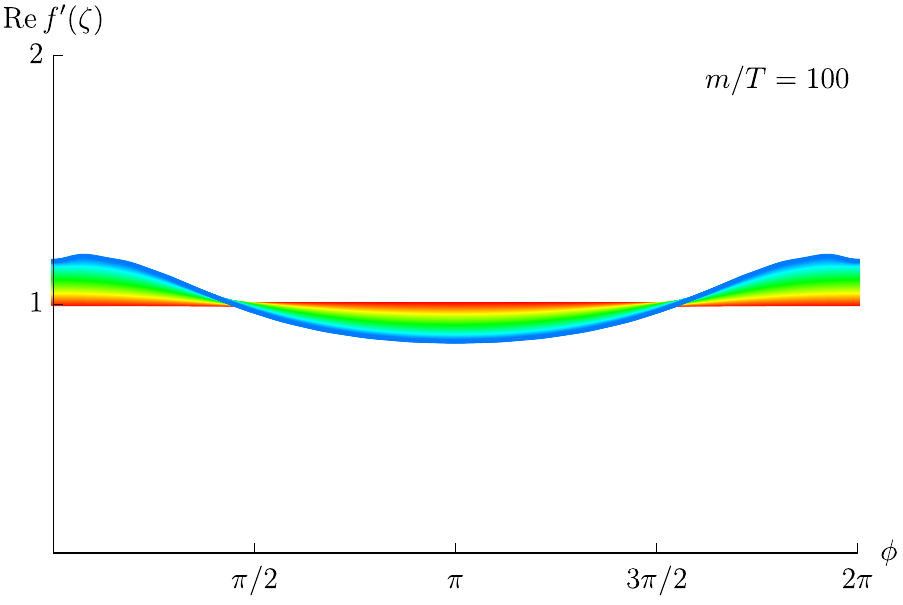}
\caption{Plot of the univalence condition $\re f'(\zeta)$, with $\zeta = |\zeta| e^{i\phi}$ and $\phi \in [0 , 2\pi]$, for energy diffusion in the linear axion model at two different values of $m/T$: $m/T = 50$ (\textbf{left panel}) and $m/T = 100$ (\textbf{right panel}). In each case, the colour gradient indicates different $|\zeta|$, from $|\zeta| = 0$ in red to $|\zeta| = 1 $ (corresponding to $|z| = R$) in blue. The function $f(\zeta)$ follows from a map of the dispersion relation $\omega(z)$ which is obtained by a polynomial fit of order $40$ to the numerically computed quasinormal mode data. \label{fig:diffusionunivalence50}}
\end{figure}

We demonstrate the univalence on the entire disk of convergence by plotting $\re f'(\zeta)$ (cf.~Eq.~\eqref{MacGregor}) for two chosen values of $m/T$ in Figure~\ref{fig:diffusionunivalence50}. The function $f(\zeta)$, which is defined as 
\begin{equation}
    f(\zeta) \equiv \frac{i \omega(R \zeta )}{DR},
\end{equation}
is then univalent on the open unit disk $|\zeta| < 1$. Here, $D$ is the (energy) diffusion constant $D_e$ discussed in Section~\ref{sec:LinearAxion_Intro} (see also Table \ref{tab1}). We observe that $\re f'(\zeta)$ behaves in a qualitative manner that is analogous to the cases of momentum diffusion (in the shear channel) in thermal theories in $d=3$ (hydrodynamic theory of M2 branes) and $d=4$ ($\CN = 4$ SYM) boundary dimensions dual to the corresponding Schwarzschild black brane geometries ($m=0$) studied in Ref.~\cite{Grozdanov:2020koi}.

\begin{figure}[h!]
\centering
\includegraphics[width=0.49\textwidth]{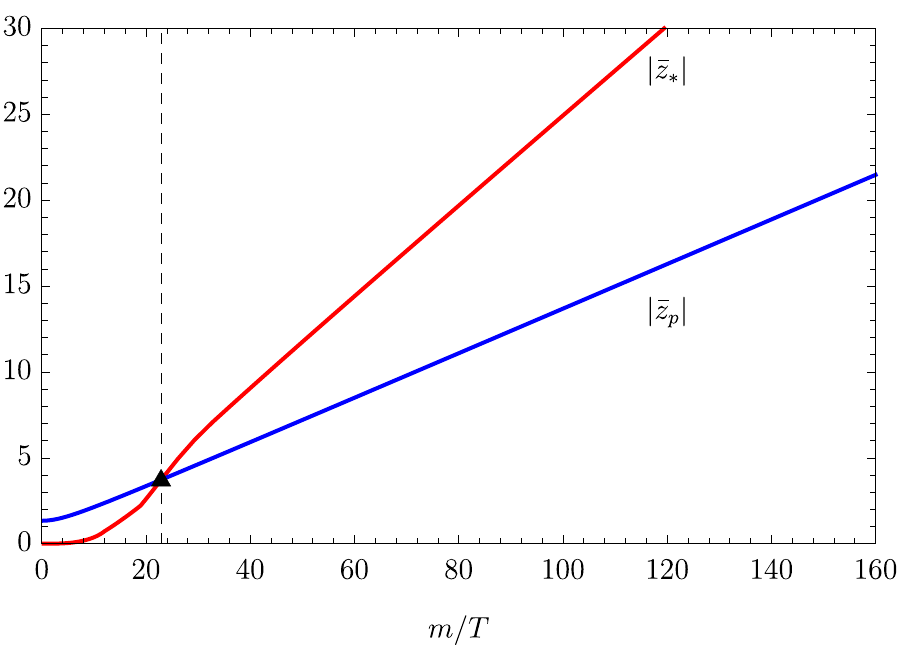}
\includegraphics[width=0.49\textwidth]{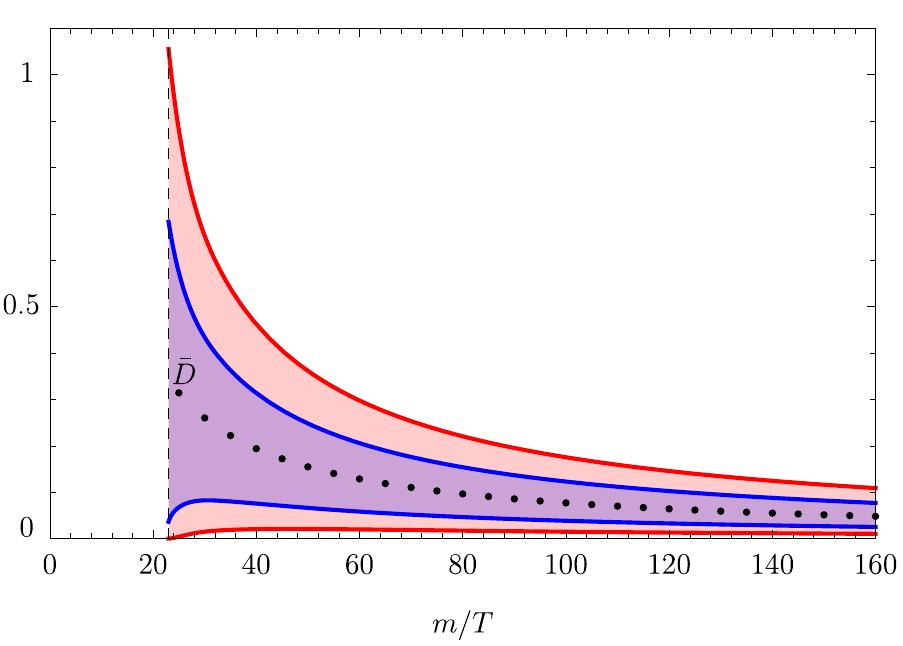}
\caption{\textbf{Left panel: } A plot of the radius of convergence $R = |\bar{z}_*|$, where $z_*$ denotes the leading critical point, and the absolute value of the location of the pole-skipping point $|\bar{z}_p|$ as a function of $m/T$. The point at which the pole-skipping point enters the convergence disk is denoted by a black triangle. Eq.~\eqref{D_PS_Growth} can only be applied to the region on the right of the vertical dashed line. \textbf{Right panel: }Bounds on energy diffusion for different values of $m/T$ using pole-skipping. The blue band indicates the logarithmic bounds from Eq.~\eqref{D_PS_Log}. For comparison, in red, we also show the looser bounds from Eq.~\eqref{D_PS_Growth}. The black dots represent the actual values of $\bar{D}$ computed numerically. \label{fig:boundforD}}
\end{figure}

Next, we compare the numerically computed values of the diffusion constant $D$ for different $m/T$ with the bounds that follow from the knowledge of the pole-skipping point~\cite{Grozdanov:2017ajz,Blake:2017ris,Blake:2018leo,Grozdanov:2018kkt}. The butterfly velocity and the Lyapunov exponent, which feature in the leading pole-skipping point, can be obtained analytically \cite{Blake:2018leo}:
\begin{equation}\label{Vb_Ly_m}
\lambda_{L}=2 \pi T = \frac{1}{2}\left(3-\frac{m^{2}}{2}\right), \qquad  v_{B}=\frac{1}{2}\sqrt{3-\frac{m^{2}}{2}}. 
\end{equation}
At the pole-skipping point $\omega_p (z_p)$, we therefore have
\begin{equation}
    \omega_p = i \lambda_L = \frac{i}{2}\left(3-\frac{m^{2}}{2}\right), \qquad z_p  = - \frac{\lambda_L^2}{v_B^2} = - \left(3- \frac{m^2}{2} \right).
\end{equation}
The values of $z_p$ are negative for all $m$ because in the longitudinal channel, $q_p$ at the (hydrodynamic) pole-skipping point is imaginary. Since $\omega(z)$ is univalent on the disk of convergence, this means, by the rescaling argument discussed in Section \ref{sec:Univalence}, that we can automatically use the tighter {\it log} bounds \eqref{GrowthMacGregor}:
\begin{equation}\label{D_PS_Log}
    \frac{\lambda_L} {R}\,\left(\ln \left[\frac{e^{-\lambda_L^2 / R v_B^2 }}{\left(1 - \frac{\lambda_L^2}{R v_B^2} \right)^2}\right]\right)^{-1} \leq D \leq     \frac{\lambda_L} {R}\left(\ln \left[e^{-\lambda_L^2 / R v_B^2 } \left(1 + \frac{\lambda_L^2}{R v_B^2} \right)^2 \right]\right)^{-1},
\end{equation}
where $\lambda_L = 2\pi T$. For comparison, the looser bounds that follow from the growth theorem \eqref{GrowthTheorem} are
\begin{equation}\label{D_PS_Growth}
    \frac{v_B^2}{\lambda_L} \left(1 - \frac{\lambda_L^2}{R v_B^2} \right)^2 \leq D \leq \frac{v_B^2}{\lambda_L} \left(1 + \frac{\lambda_L^2}{R v_B^2} \right)^2 .
\end{equation}
As noted above, the actual value of $D$ as well as the radius of convergence $R$ must be computed numerically. We present the results for different values of $m/T$ in Figure \ref{fig:boundforD}, where we use the following dimensionless quantities measured in terms of the temperature:
\begin{equation}
\bar\omega \equiv \frac{\omega}{2\pi T}, \qquad \bar z \equiv \frac{z}{(2\pi T)^2}, \qquad \bar D \equiv 2\pi T D.
\end{equation}
Note that for small $m/T$, the pole-skipping point lies outside of the chosen $U = \mathbb{D}_R$ and therefore the bounds derived above cannot be used. The locations of the pole-skipping points and the radii of convergence for different $m/T$ are depicted in the left panel of the Figure \ref{fig:boundforD}. The right panel shows the comparison between the numerically computed $D$ and the univalence bounds in Eq.~\eqref{D_PS_Log} (and \eqref{D_PS_Growth}).

It is also instructive, and possible, to enlarge the univalent region $U$ to a disk larger than the domain of convergence $\mathbb{D}_R$. To see how this works, we first note that for such cases, $\re f'(z) > 0$ no longer implies that $\re f'(\zeta) > 0$. Thus, in general, we can no longer directly apply the {\it log} bounds \eqref{GrowthMacGregor} and \eqref{MacGregorCoeffs}. Following \cite{Grozdanov:2020koi}, let us define the univalence region $U$ as the disk centred at $z = z_c$ (where $z_c$ negative and real) and with the boundary at $z = z_c \pm z_b$ (where $z_b$ is positive and real). The value of $z_b$ is directly determined by the radius of convergence of the hydrodynamic series. For positive and real $z_*$ (where $z_*$ is the leading critical point), as in our case studied here, we have $z_c + z_b = z_*$. We choose the M\"{o}bius transformation $\varphi(z)=\zeta$ to map $0 \to 0$ and $(z_c \pm z_b) \to \pm 1 $. Note that this is a somewhat different choice compared to the example shown in \cite{Grozdanov:2020koi}. For present purposes, this is the appropriate choice because the M\"{o}bius transformation maps the critical point to the boundary of the unit disk. This is required in order for the series representation of $f(\zeta)$ to remain holomorphic (and univalent) on $\mathbb{D}$. Thus,
\begin{equation}\label{Conformalmap}
    \varphi(z) = \frac{z_b z}{z_c z + z_b^2 - z_c^2}, \qquad \varphi^{-1}(\zeta) = \frac{\left(z_b^2 - z_c^2 \right)\zeta}{z_b - z_c \zeta},
\end{equation}
which gives $\partial^n_\zeta \varphi^{-1}(0) = n! (z_c)^{n-1} (z_b^2 - z_c^2) / z_b^n$. For $f(z)$ satisfying the univalence condition $\re f'(z) > 0$ in the disk $U$, then
\begin{equation}
    \re f'(z) = \re \left[\frac{\partial \varphi(z)}{\partial z} f'(\zeta) \right] = \frac{1}{z_b \left(z_b^2-z_c^2\right)}  \re \left[ \left( z_b - z_c \zeta \right)^2 f'(\zeta) \right] > 0 ,
\end{equation}
where by $f(\zeta)$ we really mean $f(z = \varphi^{-1}(\zeta))$. This means that for complex $\zeta \in \mathbb{D}$, we cannot ascertain that, generically, $\re f'(\zeta) >0$ if $\re f'(z) > 0 $. Hence, unless for some reason we could check independently that $\re f'(\zeta) > 0$, we need to use the bounds in Eqs.~\eqref{GrowthTheorem} and \eqref{deBranges}. For the diffusion constant, and with the known point $z_0$ again chosen as the pole-skipping point, we get 
\begin{align}\label{bounds_zc_ps}
    &\frac{v_B^2}{\lambda_L}\left|1 - \frac{z_c \lambda_L^2}{\left(z_b^2 - z_c^2 \right) v_B^2} \right| \left(1 - \frac{z_b \lambda_L^2}{\left| - z_c \lambda_L^2 + \left(z_b^2 - z_c^2\right) v_B^2 \right|} \right)^2 \leq D \nn 
    &\leq  \frac{v_B^2}{\lambda_L}\left|1 - \frac{z_c \lambda_L^2}{\left(z_b^2 - z_c^2 \right) v_B^2} \right| \left(1 + \frac{z_b \lambda_L^2}{\left|-  z_c \lambda_L^2 + \left(z_b^2 - z_c^2\right) v_B^2 \right|} \right)^2 .
\end{align}

\begin{figure}[h!]
\centering
 \includegraphics[width=1\textwidth]{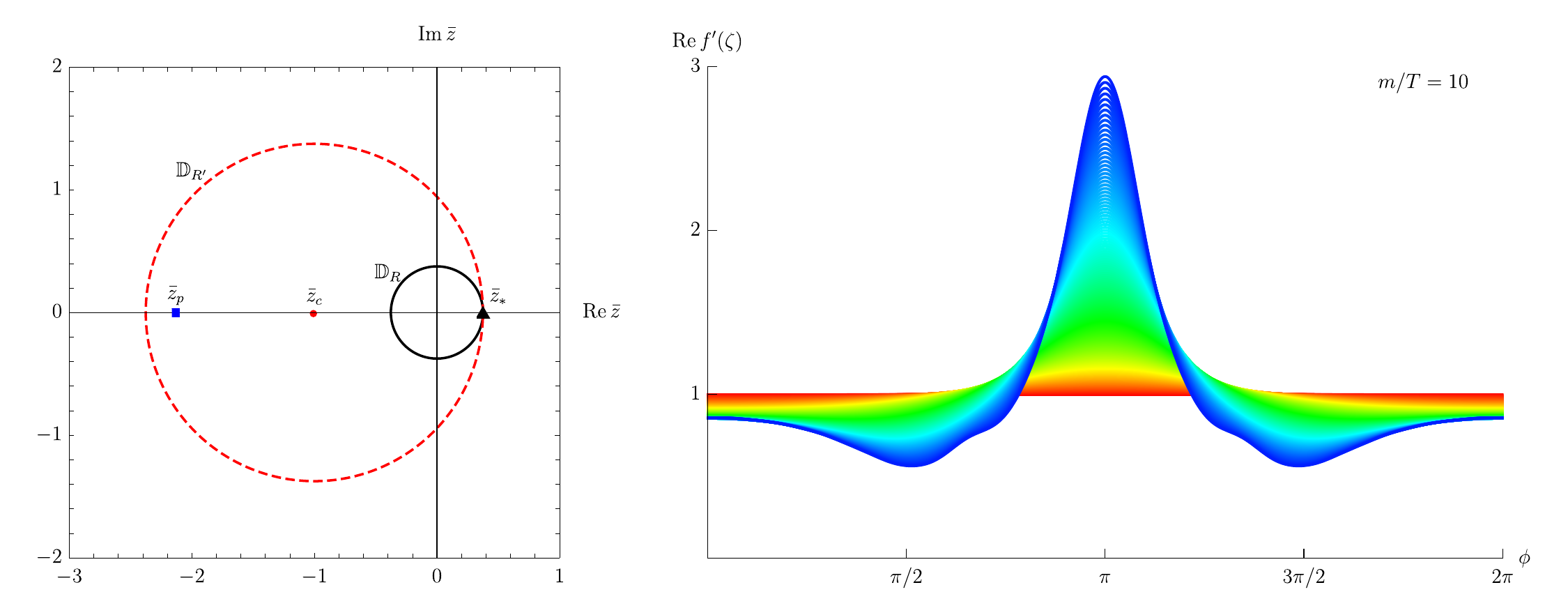}
\caption{\label{fig:univalencecondition10}\textbf{Left panel: }A plot showing the convergence disk domain $\mathbb{D}_R$ (black solid line) and an example of a larger disk $U=\mathbb{D}_{R'}$ (red dashed line) where $\omega(z)$ at $m/T = 10$ is still univalent. The centre of $U=\mathbb{D}_{R'}$ is at $\bar z_c = -1$, the location of the pole-skipping point is at $\bar z_p$ and the critical point limiting holomorphicity (convergence) is at $\bar z_* = R/(2 \pi T)^2$. \textbf{Right panel: }Here, we show that the dispersion relation is indeed univalent in $U=\mathbb{D}_{R'}$ by mapping it to the unit disk $|\zeta| < 1$ via the conformal map \eqref{Conformalmap} and plotting $\re f'(\zeta)$. As can be seen, $\re f'(\zeta) > 0$ over the entire $U \to \mathbb{D}$. Thus, in principle, the stronger {\it log} bound on diffusion \eqref{D_bound_MacG} could also be applied here. The colour scheme is the same as in Figure~\ref{fig:diffusionunivalence50}.}
\end{figure}

An example where this discussion is directly applicable can be demonstrated for values of $m/T$ with the pole-skipping point located outside of the disk of hydrodynamic convergence (see the left panel of Figure~\ref{fig:boundforD}). For concreteness, we choose $m/T = 10$. For this case, we show in Figure \ref{fig:univalencecondition10} that $\omega(z)$ remains univalent for the largest possible disk centred at $\bar z = \bar z_c = -1$. In fact, in this example, it is even still true that after using the map \eqref{Conformalmap}, $\re f'(\zeta) >0$. Hence, this again allows us to use the {\it log} version of the inequalities in~\eqref{bounds_zc_ps}: 
\begin{align}\label{bounds_zc_log}
    &\frac{z_b \lambda _L}{\left(z_b^2-z_c^2\right) \left[-\frac{z_b \lambda _L^2}{v_B^2 \left(z_b^2-z_c^2\right)-z_c \lambda _L^2}-2 \ln \left(\frac{z_b \lambda _L^2}{v_B^2 \left(z_b^2-z_c^2\right)-z_c \lambda _L^2}-1\right)\right]}\leq D \nn
    &\leq \frac{z_b \lambda _L}{\left(z_b^2-z_c^2\right) \left[2 \ln \left(\frac{z_b \lambda _L^2}{v_B^2 \left(z_b^2-z_c^2\right)-z_c \lambda _L^2}+1\right)-\frac{z_b \lambda _L^2}{v_B^2 \left(z_b^2-z_c^2\right)-z_c \lambda _L^2}\right]}.
\end{align}
For this specific example (see Figure~\ref{fig:univalencecondition10}), we find the bounds on the diffusivity that now follow from~\eqref{bounds_zc_log} to be
\begin{equation}
0.26\leq \bar{D}\leq 3.99, 
\end{equation}
while its numerically computed value is  $\bar{D} = 0.86$.

It is important to note that the choice of $z_c$ (inside $U$) is of fundamental importance to the derivation of these bounds. However, at present, a robust mechanism for making the choice that would lead to `optimal' bounds is missing and will have to be understood in the future.

In some examples, it may even be possible to take $-z_c$ all the way to infinity. Then, since $z_b + z_c = z_* = R$ remains fixed, $z_b^2 - z_c^2 = (z_b + z_c)(z_b - z_c) \to - 2 z_c R$ and $z_b / z_c \to -1$. Hence, in this limit, the inequalities from Eq.~\eqref{bounds_zc_ps} give
\begin{align}
    \frac{2 R v_B^4}{\lambda_L \left(\lambda_L^2 + 2 R v_B^2\right)} \leq D \leq \frac{2 \left( \lambda_L^2 + R v_B^2 \right)^2 }{\lambda_L R \left(\lambda_L^2 + 2 R v_B^2  \right)}  .
\end{align}
For the linear axion model, if the limit $z_c \to - \infty$ is permitted, thus,
\begin{align}\label{AxionBoundExtDisk}
    \frac{2R}{6-m^2+4 R} \leq D \leq  \frac{\left(6 - m^2 + 2 R\right)^2}{2 R \left(6-m^2 +4 R\right)}.
\end{align}

An example when this is possible and Eq.~\eqref{AxionBoundExtDisk} can be tested is the self-dual point at $m=\sqrt{2}$ \cite{Davison:2014lua}. There, the hydrodynamic dispersion relation is given simply by the expression
\begin{equation}
    \bar\omega(\bar z) =  - \frac{1}{2} i \left(1 - \sqrt{1 - 4 \bar z } \right), 
\end{equation}
with the branch point at $z_* = R = 1/4$. The bounds on diffusion for an initially chosen (infinite) univalent disk \eqref{AxionBoundExtDisk} are then
\begin{equation}
    0.1 \leq \bar{D} \leq 8.1.
\end{equation}
This result is weaker than the bounds derived by using the Koebe function map in Ref.~\cite{Grozdanov:2020koi}, which gave
\begin{equation}
    0.5 \leq \bar{D} \leq 4.5.
\end{equation}
This is not unexpected as the Koebe function maps the entire complex plane (without a semi-infinite branch cut line) to the final unit disk $\mathbb{D}$. We note that in this case, the actual value of the diffusivity is $\bar{D}=1$.

\begin{figure}[h!]
\centering
 \includegraphics[width=0.65\textwidth]{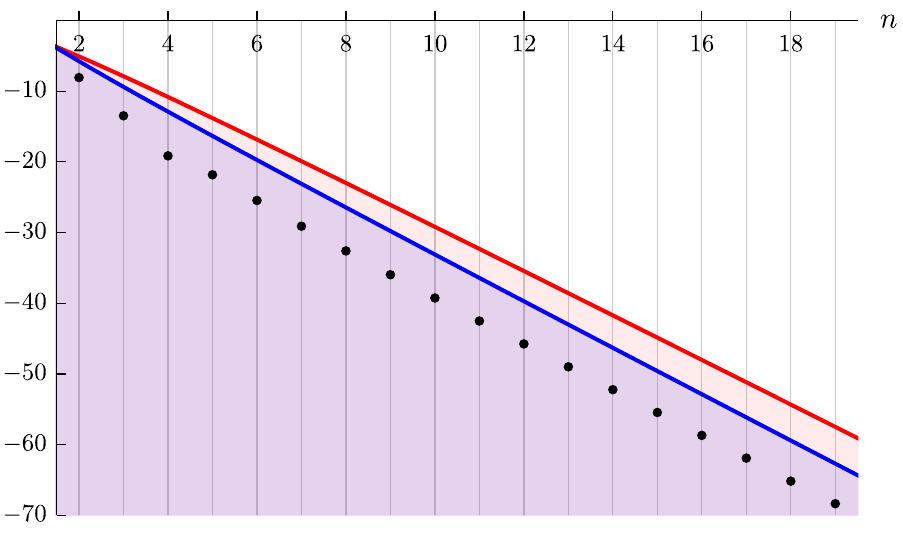}
\caption{\label{fig:higherorderbounds}The univalence bounds on the series coefficients $\bar{c}_{n\geq2}$ for the hydrodynamic energy diffusion mode in the linear axion model with $m/T=100$. Black dots represent the logarithms of the absolute values of the numerically computed coefficients, i.e.,~$\ln{|\bar{c}_n|}$. The red line represents the logarithm of the bounds set by the de Branges theorem in Eq.~\eqref{deBrangesforcn} (i.e.,~$ \ln n \bar{D}/\bar R^{n-1} $) while the blue line corresponds to the bounds given by the logarithms of the stronger {\it log} bounds in Eq.~\eqref{MacGregorCoeffsforcn} (i.e.,~$ \ln 2 \bar{D}/ n \bar R^{n-1} $).}
\end{figure}

Next, we turn our attention to an example of bounds applied to the higher-order coefficients $c_n$ in the diffusive dispersion relation \eqref{WDiff}. We again take $m/T = 100$ and use the convergence disk $\mathbb{D}_R$ as the domain of univalence $U = \mathbb{D}_R$ (note that $z_b = R$ and $z_c = 0$). We refer the reader to the right panel of Figure~\ref{fig:diffusionunivalence50} where we showed that the diffusive dispersion relation was indeed univalent in $U$. Because $\re f'(\zeta) > 0$, we can directly apply Eq.~\eqref{MacGregorCoeffs}, although for comparison, we will also consider the looser Eq.~\eqref{deBranges} based on the de Branges theorem. Using Eq.~\eqref{deBranges}, we find 
\begin{equation}
\label{deBrangesforcn}
|\bar{c}_{n\geq 2}| \leq \frac{n \bar{D}}{\bar R^{n-1}}.
\end{equation}
Eq.~\eqref{MacGregorCoeffs} instead implies the stronger {\it log} bounds 
\begin{equation}
\label{MacGregorCoeffsforcn}
|\bar{c}_{n \geq 2}| \leq \frac{2 \bar{D}}{n \bar R^{n-1}},
\end{equation}  
where we have used the dimensionless coefficients $\bar c_n \equiv (2\pi T)^{2n-1} c_n$. $\bar R$ is the dimensionless radius of convergence $\bar R \equiv R / (2\pi T)^2$. Numerically, we find $\bar{D} = 0.077$ and $\bar R = 24.908$. We compare these bounds with the numerically computed coefficients $\bar c_n$ in Figure~\ref{fig:higherorderbounds}. The plotted absolute values of lower and upper bounds are equal to each other.  

\section{Holographic solids with spontaneously broken translations: \\ diffusion and sound}\label{sec:solids}

We turn now to the second subclass of holographic axion models in which translations are broken spontaneously in the boundary field theory, rather than explicitly. As discussed in Section~\ref{sec:Solids_Intro}, we consider a specific model with $V(X,Z)=X^3$. The temperature $T$ and the parameter $m/T$ describing the relative importance between the `amount' of broken translations and temperature are now given by
\begin{equation}
T= \frac{6-m^6}{8 \pi} , \qquad \frac{m}{T} = \frac{8 \pi m}{6- m^6}\,. 
\end{equation}
Physically, in this situation, $m/T$ represents the amount of `solidity' of the boundary field theory --- the rigidity of the dual `solid'. To obtain the quasinormal mode frequencies numerically and thereby determine the spectrum of the boundary field theory, we employ a pseudo-spectral method. All equations and numerical methods used in this Section are explained in detail in Refs.~\cite{Ammon:2019apj, Baggioli:2019abx, Grieninger:2020wsb}.

In this theory, the location of the hydrodynamic pole-skipping point in the longitudinal channel was analysed in \cite{Jeong:2021zhz}:
\begin{equation}\label{PS_Solid1}
\lambda_{L}=2 \pi T = \frac{1}{2}\left(3-\frac{m^{6}}{2}\right), \qquad  v_{B}=\frac{1}{2}\sqrt{3-\frac{m^{6}}{2}}. 
\end{equation}
At the pole-skipping point $\omega_p (z_p)$, where here, $z \equiv q = \sqrt{\q^2}$ instead of $q^2$, we therefore have
\begin{equation}\label{PS_Solid2}
    \omega_p = i \lambda_L = \frac{i}{2}\left(3-\frac{m^{6}}{2}\right), \qquad z_p  = i \frac{\lambda_L}{v_B} = i \sqrt{ 3- \frac{m^6}{2} }.
\end{equation}
We recall that in the longitudinal channel, the hydrodynamic spectrum contains a pair of sound modes and a diffusive mode. In the transverse channel, where we will not need to know the locations of the pole-skipping points, the hydrodynamic spectrum only contains a pair of sound modes (see Table~\ref{tab1}). Due to the coexistence and the simultaneous discussions of diffusion and sound, in this section, we will always use the variable $z$ defined as $z \equiv q$. 

The regions $U$ that will be considered here will always be disks $\mathbb{D}_r$ of some radius $r$ centred at $z=0$. In most cases, both for diffusive and sound modes, we will see that the entire disk of convergence is also univalent, i.e., that $r = R$. Only in one of the considered examples will the radius $r$ be set by the local univalence breaking condition \eqref{local_cond} of vanishing group velocity. In that case, we will have $r = |z_g| < R$, with $z_g$ imaginary, which is analogous to the behaviour of the translationally invariant $\CN =4 $ SYM theory studied in Ref.~\cite{Grozdanov:2020koi}.

Firstly, we consider the longitudinal channel. We focus on three concrete values of $m/T=\{3.31,6.18,8.37\}$, which we have found to be sufficient in order for us to be able to qualitatively illustrate the univalence properties in this theory. We label these cases as I${}_L$, II${}_L$ and III${}_L$, respectively. To analyse the univalence properties of the sound and the diffusive dispersion relations in this channel, we numerically compute their respective convergence radii $R_s$ and $R_d$ in terms of the $z = q$ variable. The dimensionless counterparts are defined as $\bar R_s \equiv R_s / (2\pi T)$, $\bar R_d = R_d / (2\pi T)$. Intriguingly, in Case I${}_L$ with $m/T = 3.31$, we find that the radii of convergence of the two series are equal $R_d = R_s$. In other words, the leading (lowest) critical point for both types of hydrodynamic modes is given by a single pair of critical points where the two hydrodynamic modes collide and limit each other's convergence. This phenomenon appears to be rather robust for `small' values of $m/T$ and persists all the way to very small $m/T \rightarrow 0$, where we observe that the collision between the two hydrodynamic modes tends towards a purely imaginary wavevector value $\bar z=i/(2\pi T)$. To the best of our knowledge, this has not been observed in other holographic models. We discuss the details of this collision in Appendix~\ref{appendix:Coll}. In each of the three cases, we also numerically compute the speed of sound $v_s$ and the diffusion constant $D$, with dimensionless $\bar D \equiv 2\pi T D$. The pole-skipping momentum is obtained analytically from Eq.~\eqref{PS_Solid2}. Its dimensionless absolute value is defined as $|\bar z_p| \equiv |z_p / (2\pi T)|$. We note that in Cases I${}_L$ and II${}_L$, the pole-skipping point lies outside of the convergence disk $\mathbb{D}_R$, while in Case III${}_L$, it lies inside and can therefore be easily used in the chaos-type univalence bounds. 

\begin{figure}[h!]\centering
\includegraphics[width=0.48\textwidth]{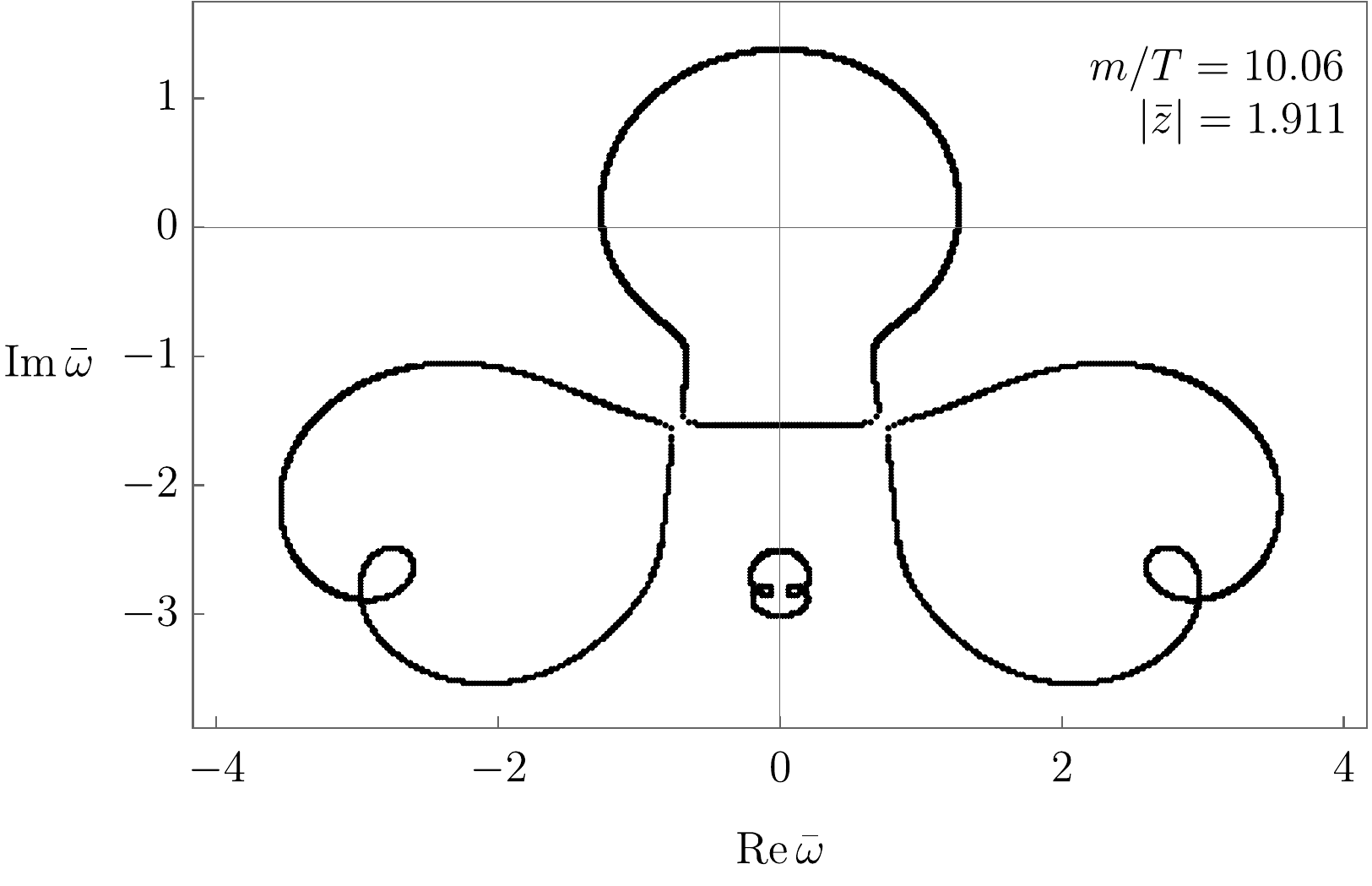} \includegraphics[width=0.48\textwidth]{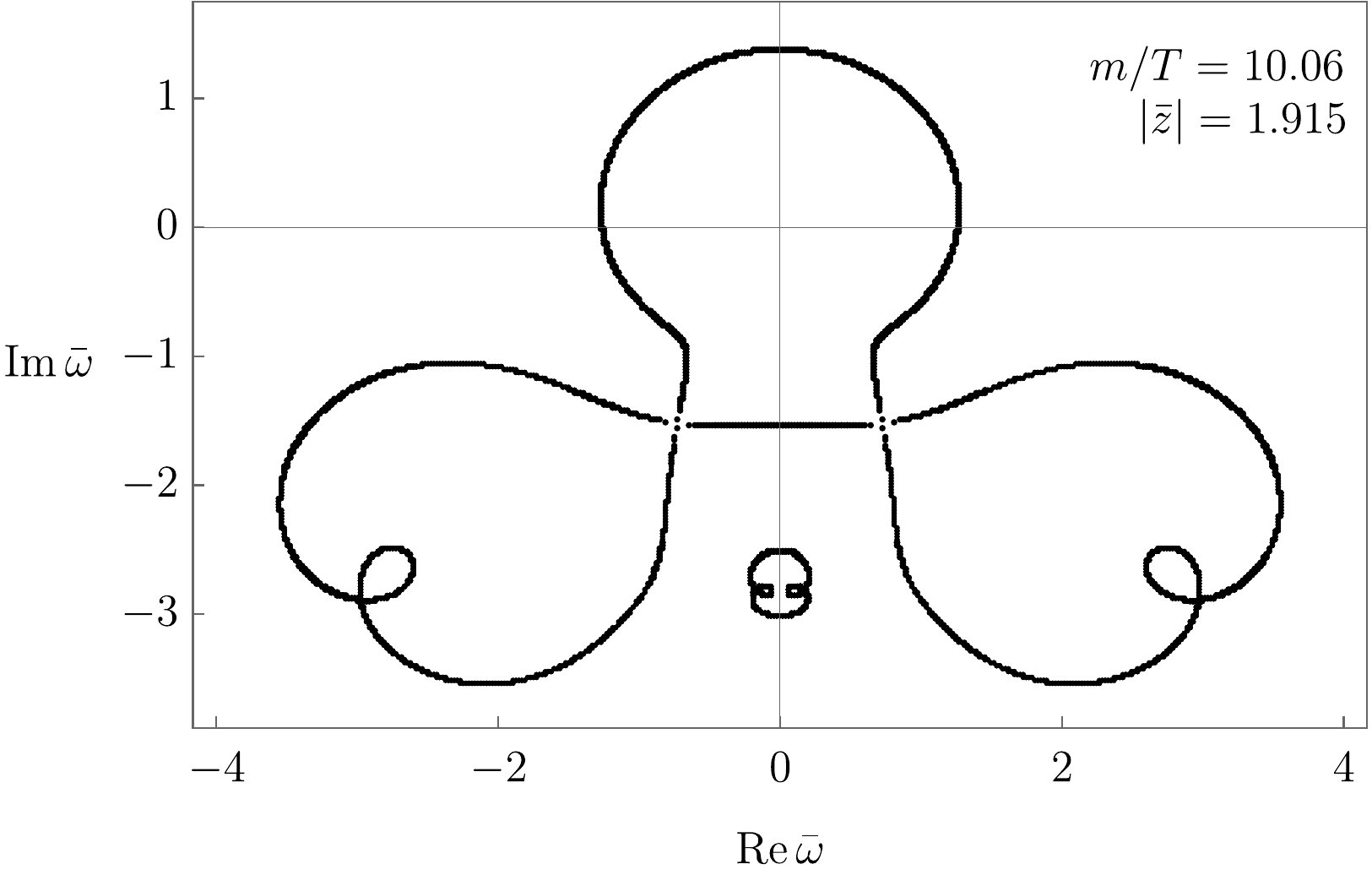}
\caption{Plots of $\bar \omega ( \bar z)$ for $\bar z = |\bar z| e^{i\phi}$, with $\phi \in [0,2\pi]$, before and approximately at the leading collision of the pair of transverse sound modes with a pair of gapped modes in the holographic solid model for $m/T = 10.06$. This collision sets the radius of convergence to $\bar R_s = 1.92$.}
\label{SolidTransCollision}
\end{figure}

In the transverse channel, which contains a pair of sound modes, we choose two representative values of $m/T = \{3.31, 10.06\}$ and label these cases as I${}_T$, II${}_T$. Here too, we numerically compute the radius of convergence of the hydrodynamic sound series $\bar R_s$ and the speed of sound $v_s$. Since there is only one type of hydrodynamic excitation present in this channel, the collision that limits convergence of the hydrodynamic series is of the most common type: a collision between a hydrodynamic and a gapped mode. We show an example of such a collision in Figure~\ref{SolidTransCollision}. Moreover, as will be discussed below, in Case I${}_T$, we also compute the (imaginary) value $\bar z  = z /  (2 \pi T)$, denoted by $z_g$, at which the group velocity of the mode $v_g$ vanishes and univalence is violated locally. This is the only case in which $z_g$ exists within the disk of convergence $\mathbb{D}_R$. Other quantities discussed in the longitudinal channel will play no role in the analysis of the transverse channel.

Explicit values of all relevant quantities in the longitudinal and the transverse channels are collected in Table~\ref{tab2}.

\begin{table}[h!]
\centering
\begin{tabular}{ |p{1.8cm}||p{0.8cm}|p{0.8cm}|p{0.8cm}|p{0.8cm}|p{0.8cm}|p{0.8cm}|p{1cm}|}
\hline
\multicolumn{8}{|c|}{\cellcolor{blue!08} \textbf{Holographic solid}} \\
\hline
\hline
\cellcolor{gray!10}  & \cellcolor{gray!10} $m/T$  & \cellcolor{gray!10} $\bar{R}_s$ & \cellcolor{gray!10} $\bar{R}_d$ &  \cellcolor{gray!10} $|\bar{z}_p|$ & \cellcolor{gray!10} $|\bar z_g|$ & \cellcolor{gray!10} $\bar D$ & \cellcolor{gray!10} $v_s$ \\
\hline
\hline
\multicolumn{8}{|c|}{\cellcolor{green!08} \textbf{Longitudinal hydrodynamic modes}} \\
\hline
\hline
Case I${}_L$ & 3.31 & 0.77& 0.77  & 1.17 & \xmark & 1.31  & 0.75 \\ 
 \hline
Case II${}_L$ & 6.18 & 1.09  & 1.31 & 1.35 &\xmark & 0.71 & 0.915 \\
  \hline
Case III${}_L$ & 8.37 & 1.36  & 1.78  & 1.51 &\xmark & 0.52  & 0.96  \\
  \hline
  \hline
\multicolumn{8}{|c|}{\cellcolor{green!08} \textbf{Transverse hydrodynamic modes}} \\
\hline
\hline
Case I${}_T$ & 3.31 & 1.16& \xmark  & \xmark & 0.26 & \xmark  &0.25  \\ 
 \hline
Case II${}_T$ & 10.06 &  1.92 & \xmark & \xmark & \xmark & \xmark &0.67  \\
\hline
\end{tabular}
\caption{Relevant quantities for the analysis of univalence in the longitudinal and the transverse channels of the holographic solid model: convergence radii of the hydrodynamic sound and diffusive series $\bar{R}_s$ and $\bar{R}_d$, the absolute value of the pole-skipping wavevector $|\bar{z}_p|$, the absolute value of the wavevector at which the group velocity of the mode vanishes $|\bar{z}_g|$, the diffusion constant $\bar D$ and the speed of sound $v_s$. All quantities (except the already dimensionless $v_s$) are stated in their dimensionless form, in units of $2\pi T$. The symbol \xmark $\,$ indicates that the corresponding value is not well-defined, does not exist or that it is irrelevant for the present analysis.}
\label{tab2}
\end{table}

\subsection{Univalence properties in the longitudinal channel}\label{sec:Solid_Long}

\begin{figure}[h!]
\centering
 \includegraphics[width=0.9\textwidth]{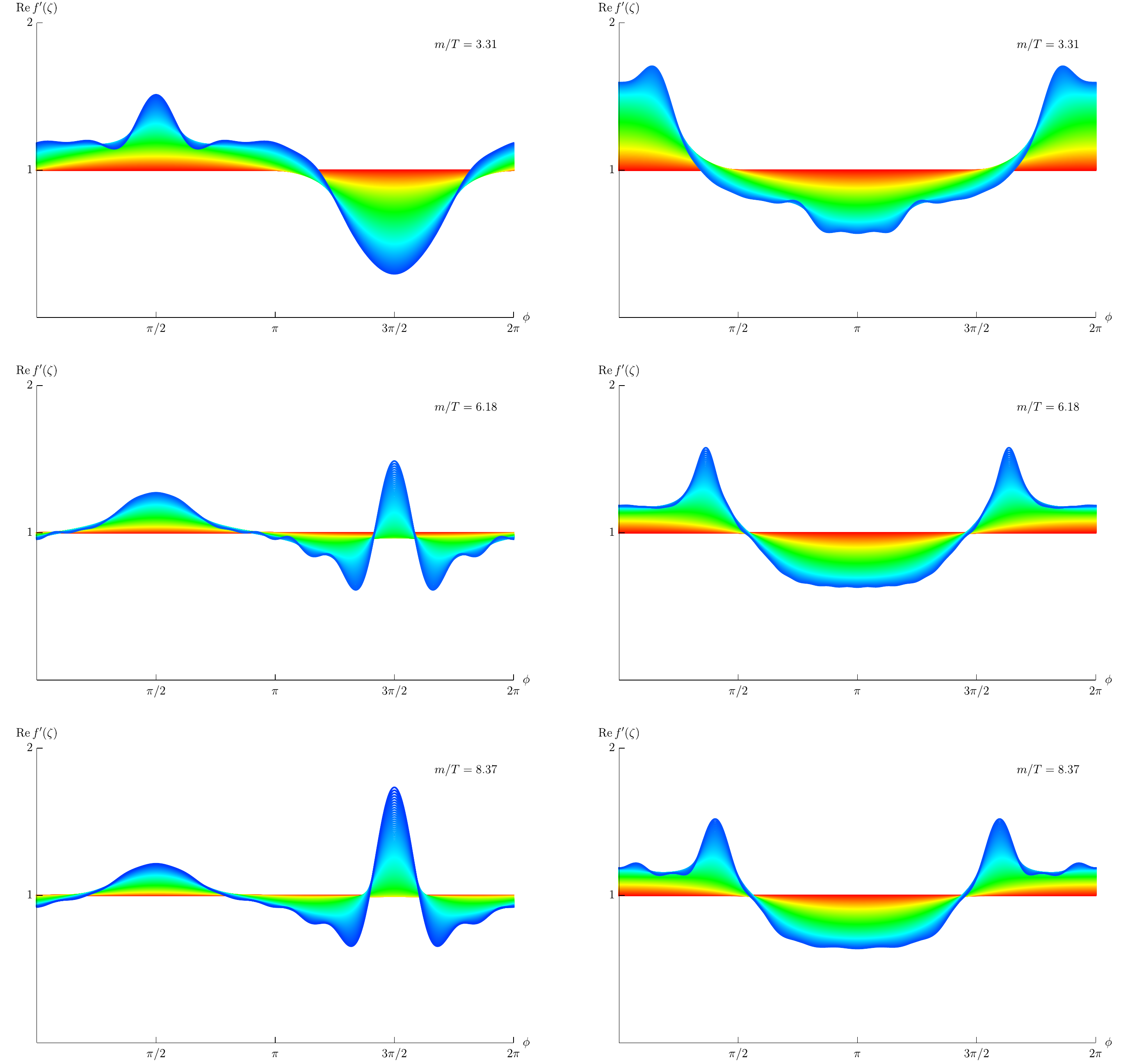}
\caption{\label{fig:UnivalenceLong} We show that in all three cases considered in the longitudinal channel, sound modes (shown in the {\bf left panels}) and diffusive modes (shown in the {\bf right panels}) are univalent on the entire disks of convergence $U = \mathbb{D}_R$. All plots are made for $\zeta = |\zeta| e^{i\phi}$ and $\phi \in [0 , 2\pi]$, with $|\zeta|$ ranging from $0$ (in red) to $0.85$ (in blue). The oscillations that are visible in the plots are a numerical artefact of the interpolations (fits) of the dispersion relations and are not physical. While the numerical uncertainties grow for larger $|\zeta|$ than those plotted here, we are confident that the dispersion relations remain univalent all the way to $|\zeta| \to1$.}
 \end{figure}

In all three cases that we consider in the longitudinal channel (I${}_L$, II${}_L$ and III${}_L$), the diffusive mode has a dispersion relation that is univalent within the entire disk of convergence $|z| < R_d$. This behaviour qualitatively matches that of the longitudinal energy diffusion mode in the linear axion model with explicit symmetry breaking (cf.~Section~\ref{sec:Diff}) and also that of the transverse momentum diffusion mode in thermal CFTs without translational symmetry breaking considered in Ref.~\cite{Grozdanov:2020koi}. Interestingly, the same is true also for the sound modes in this channel. This is unlike what was observed in translationally-invariant thermal CFTs in \cite{Grozdanov:2020koi}, where $|z_g| < R$ and therefore, univalence was broken locally within the (holomorphic) disk of convergence. Here, in all three cases, no point $z_g$ exists inside $|z| < R$ where $v_g = 0$. We demonstrate the univalence of all six cases (three cases of diffusion and three of sound) in Figure~\ref{fig:UnivalenceLong} by plotting $\re f'(\zeta) $ for $|\zeta| < 1$ after mapping $U = \mathbb{D}_R \to \mathbb{D}$. 

We can now also directly test the univalence bounds. For simplicity, and for the ability to compare this model with the linear axion model discussed in Section \ref{sec:Diff}, we will only consider the chaotic pole-skipping bounds on diffusion. In particular, we will only look at Case III${}_L$, which contains the pole-skipping point within the convergence disk (cf.~Table \ref{tab2}). Using the {\it log} bounds from Eq.~\eqref{D_PS_Log} then gives us the following lower and upper bounds on the dimensionless diffusion constant:
\begin{equation}
    0.23\leq  \bar{D} \leq 0.74 ,
\end{equation}
which is satisfied by the numerical value of the dimensionless diffusion constant $\bar{D}=0.52$. One may notice that the bounds appear extremely tight, in fact, tighter than those derived on energy diffusion in Section \ref{sec:Diff}. Finally, we can also test the {\it log} bounds from Eq.~\eqref{MacGregorCoeffsforcn} on the higher-order coefficients. We find that the values satisfy the bounds. Qualitatively, the behaviour of the coefficients in comparison to the bounds is analogous to that shown in Figure~\ref{fig:higherorderbounds}, so for conciseness, we do not display the plot here.

\subsection{Univalence properties in the transverse channel}\label{sec:Solid_Trans}

\begin{figure}[h!]
\centering
 \includegraphics[width=0.49\textwidth]{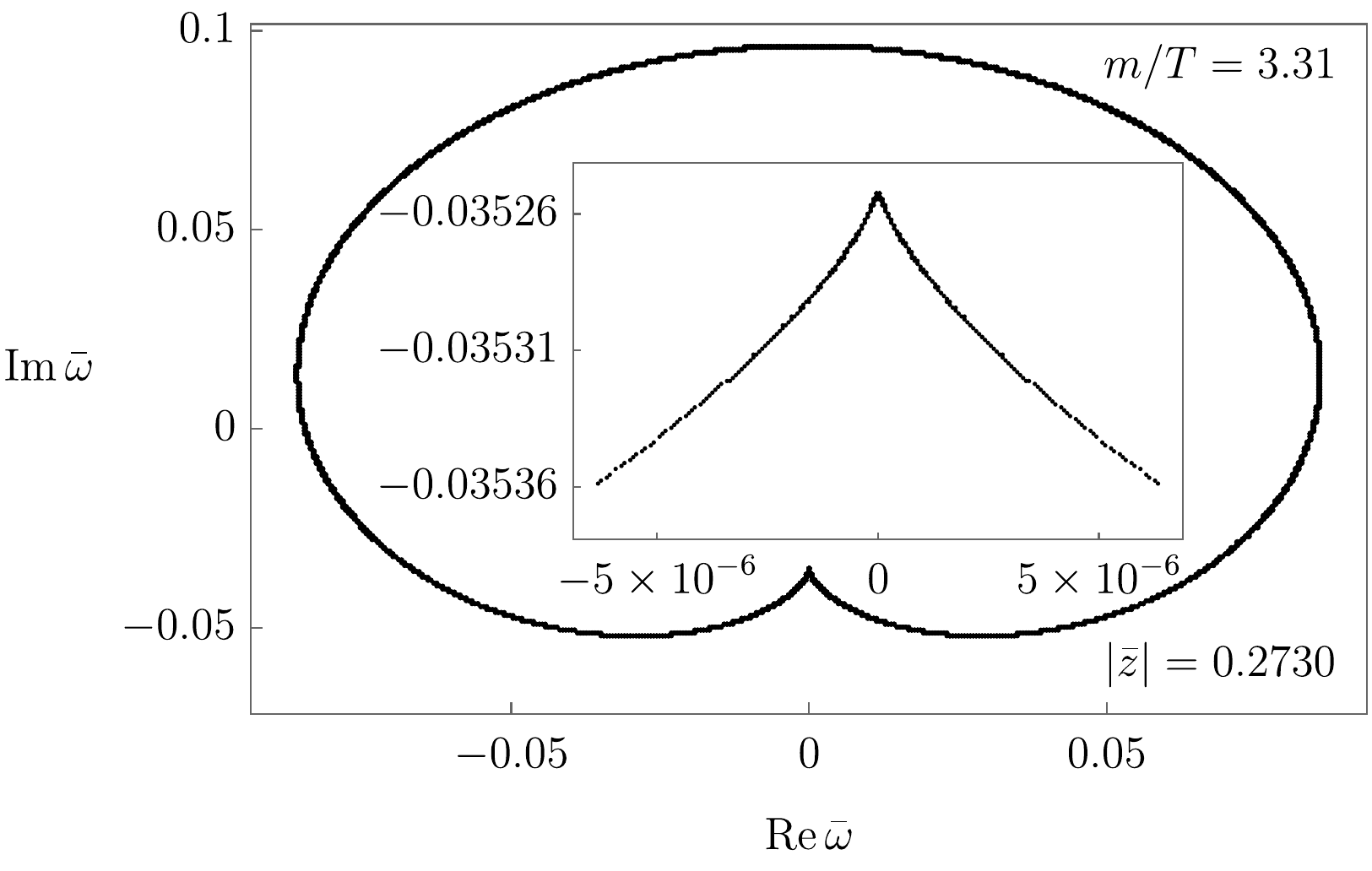} \includegraphics[width=0.49\textwidth]{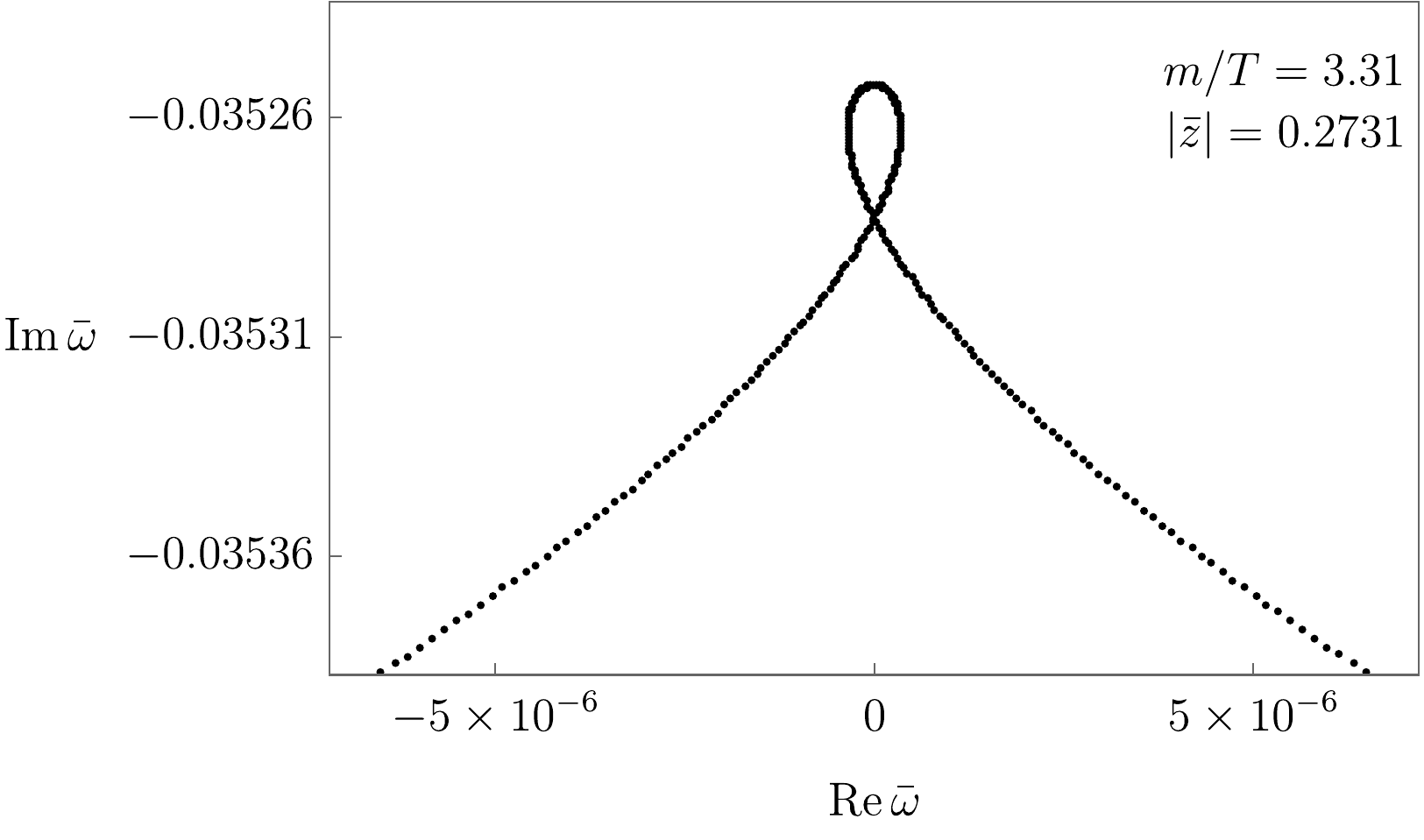}
\caption{\label{selfself}The formation of a cusp and eventual self-intersection for the transverse sound mode in Case I${}_T$ at $m/T = 3.31$. Here, the cusp of the curve $\omega(z)$ forms inside the convergence disk at $|\bar z| = 0.27 < \bar{R}_s = 1.16$. See Appendix~\ref{appendix:R}. The inlay in the left panel shows the zoomed-in version of the cusp.}
\end{figure}

As in all of the cases before, we also verify the univalence of the sound modes in the transverse channel. The two cases considered here qualitatively interpolate between the behaviour in the longitudinal channel where the sound dispersion relation was univalent on the entire disks of convergence ($U = \mathbb{D}_R$ in Case II${}_T$) to the behaviour seen for translationally invariant examples in Ref.~\cite{Grozdanov:2020koi} with $U$ determined by the local univalence breaking condition ($U = \mathbb{D}_{|z_g|}$ in Case I${}_T$). We show the formation of a cusp followed by a self-intersection in Figure~\ref{selfself}. The actual value of $\bar{z}$ at the cusp is $\bar{z}_g = 0.27 i$, which is again purely imaginary, as in the $\CN =  4$ SYM theory. For a further discussion of this fact, we refer the reader to Appendix~\ref{appendix:Cusps}. We show the behaviour of the radius of convergence as a function of $m/T$ in Appendix~\ref{appendix:R}.

For conciseness, here, we do not show any bounds on coefficients for these two specific cases as their analysis directly follows all of the cases discussed above. The only property that we investigate more closely is the speeds of sound modes in relation to the conformal bound \cite{Hohler:2009tv,Cherman:2009tw}. This is done in the next section.

\subsection{The conformal bound on the speed of sound}\label{sec:Conf}

More than a decade ago, Refs.~\cite{Hohler:2009tv,Cherman:2009tw} conjectured the possibility that certain quantum field theories have speeds of sound bounded from above by the conformal value $v_c$ defined in Eq.~\eqref{TsoundSpeed}, which is fixed by the tracelessness of the ideal hydrodynamic energy-momentum tensor. The proposed bounds are therefore
\begin{equation}\label{ConformalBound}
0 \leq v_s \leq v_c = \sqrt{\frac{1}{d-1}},
\end{equation}
where $d$ is the number of spacetime dimensions, in our case, $d=3$. Despite the numerous subsequent studies of examples that satisfy this bound, as well as of many violations (see e.g.~\cite{Kulaxizi:2008jx,Bedaque:2014sqa,Hoyos:2016cob,Anabalon:2017eri,Yang:2017oer,Ecker:2017fyh,Annala:2019puf,Ishii:2019gta,Baggioli:2020ljz}), the utility, the validity, the generality and the meaning of this statement remain unclear. Moreover, any fundamental underlying origin for the existence of this bound also remains unknown. 

Without a resolution to any of these problems, or a general proof for specific classes of quantum field theories, Ref.~\cite{Grozdanov:2020koi} derived a sufficient univalence condition on the sound dispersion relation that puts a specific theory into such a class of theories satisfying the conformal bound \eqref{ConformalBound}. As discussed in Section~\ref{sec:Solids_Intro}, for holographic solids, the speed of the transverse sound mode always satisfies the bound \eqref{ConformalBound} while the speed of the longitudinal sound mode always violates it (see Eqs.~\eqref{TsoundSpeed} and \eqref{LsoundSpeed}). For this reason, holographic solids are an ideal class of models to test various claims about the bound, including those made by appeal to univalence.

More precisely, what Ref.~\cite{Grozdanov:2020koi} argued was the following. For a sound dispersion relation, there always exists a finite disk $z \in U$ centred at $z=0$ where $\omega(z)$ is univalent. In examples considered thus far, it appears that this disk is either limited by the radius of convergence $R$ (a critical point of the spectral curve) or by the local univalence violation at $z_g$. The growth theorem based univalence bounds on the speed of sound are then (the analogue of the diffusion bound in Eq.~\eqref{D_bound_Uni}):
\begin{equation}\label{v1}
\frac{|\omega_0| \left(1 - |\zeta_0| \right)^2}{|\zeta_0| \left| \partial_\zeta \varphi^{-1}(0)\right| } \leq v_s \leq \frac{|\omega_0| \left(1 + |\zeta_0| \right)^2}{|\zeta_0| \left| \partial_\zeta \varphi^{-1}(0)\right| } .
\end{equation}
Since we will work with maps from disk to disk (rescalings), we could in principle use the {\it log} bounds, but we will stick to the standard growth theorem as it is more convenient for present purposes. One can now devise an identity between the conformal map $\varphi$, the number of dimensions $d$ and a known point of the dispersion relation $\omega_0(z_0)$ that leads to the conformal bound \eqref{ConformalBound}. Firstly, to guarantee that the lower bound is zero, we take $\zeta_0 = \varphi(z_0)$ that are at the boundary of the unit disk $\mathbb{D}$ (or infinitesimally close to it). That is, $|\zeta_0| \to 1$ or, rather, $\zeta_0 \to e^{i \phi}$, so that Eq.~\eqref{v1} becomes
\begin{equation}\label{v2}
0 \leq v_s \leq \frac{4 \left|\omega_0 (z_0) \right| }{\left| \partial_\zeta \varphi^{-1}(0)\right| } ,
\end{equation}
where $z_0 = \varphi^{-1}(e^{i\phi_0})$, with $\phi_0$ a real number between $0$ and $2\pi$. Finally, matching the upper bound to the conformal speed of sound, we arrive at the condition:
\begin{equation}\label{con1}
    |\partial_\zeta \varphi^{-1}(0)|=4 |\omega_0(z_0)|\sqrt{d-1} = 4  \left|\omega_0 \! \left( \varphi^{-1}(e^{i\phi_0})\right)\right| \sqrt{d-1},
\end{equation}
where, in our case, $d=3$. This means that so long as an $\omega_0$ at $z_0$ (or $\phi_0$) exists that satisfies Eq.~\eqref{con1}, then the speed of sound will be within the conformal bound interval. This is therefore a sufficient condition for the conformal bound \cite{Grozdanov:2020koi}. Note, however, that this is not a necessary condition, which means that if the above relation is not satisfied for a particular choice of $U$, this does not automatically imply that the conformal bound is violated. 

In our specific examples, we can further simplify these expressions by using that fact that we choose $\varphi$ to map a univalent disk $U = \mathbb{D}_r$, with either $r = R$ or $r = |z_g|$ to a unit disk $\mathbb{D}$. Therefore, $\zeta = \varphi(z) = z / r$ and $z = \varphi^{-1}(\zeta) = r \zeta$, and hence, the condition \eqref{con1} becomes
\begin{equation}\label{con2}
1 - \frac{4 \sqrt{2}}{r} \left| \omega_0 (r e^{i\phi_0}) \right| = 0 .
\end{equation}
In other words, if we can find $\phi_0$ along the circle at the edge of the univalence disk $U = \mathbb{D}_r$ such that Eq.~\eqref{con2} is satisfied, then $v_s$ in that theory will satisfy the conformal bound \eqref{ConformalBound}. 

\begin{figure}[h]
    \centering
    \includegraphics[width=0.45\linewidth]{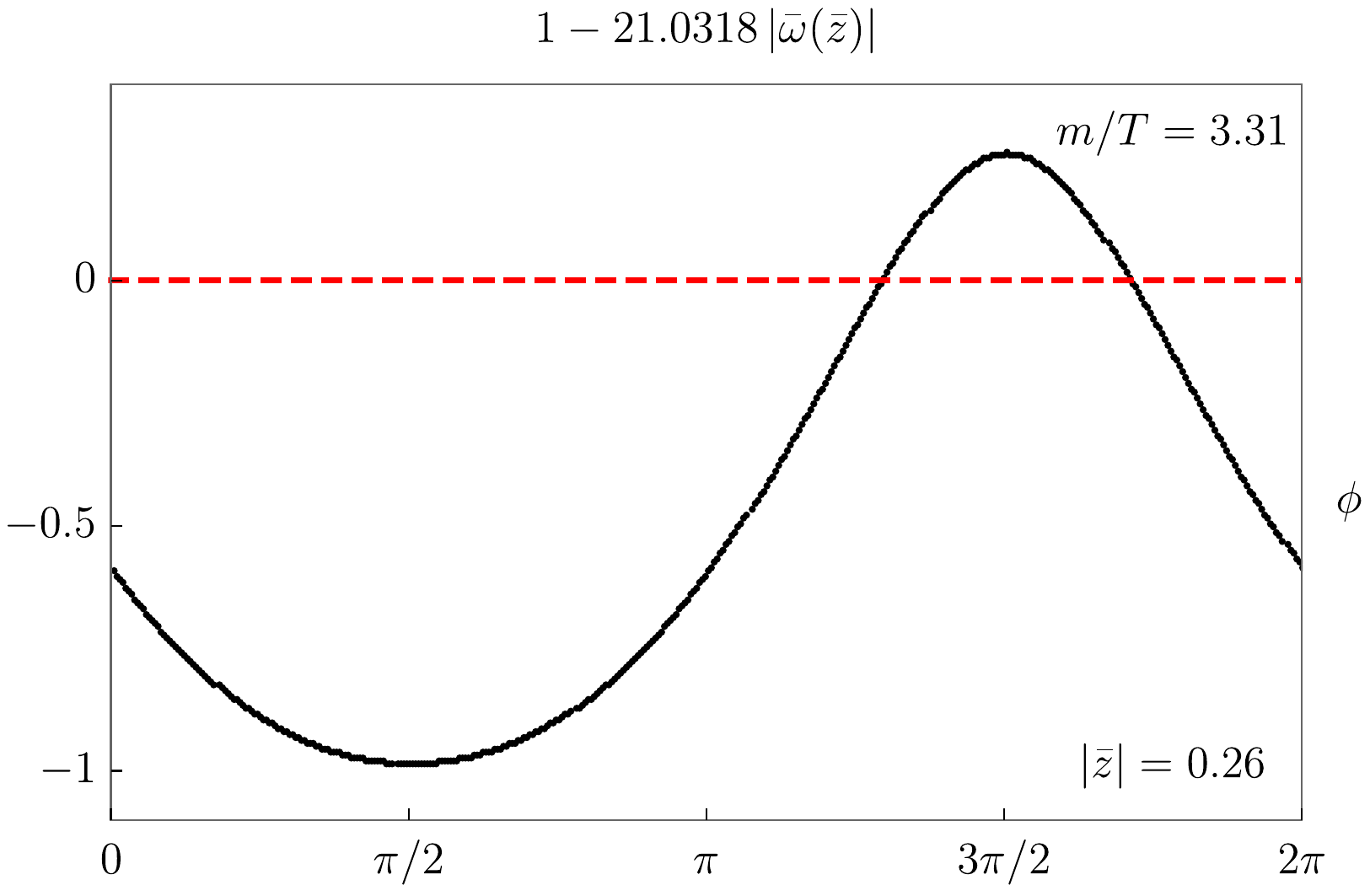} \qquad 
     \includegraphics[width=0.45\linewidth]{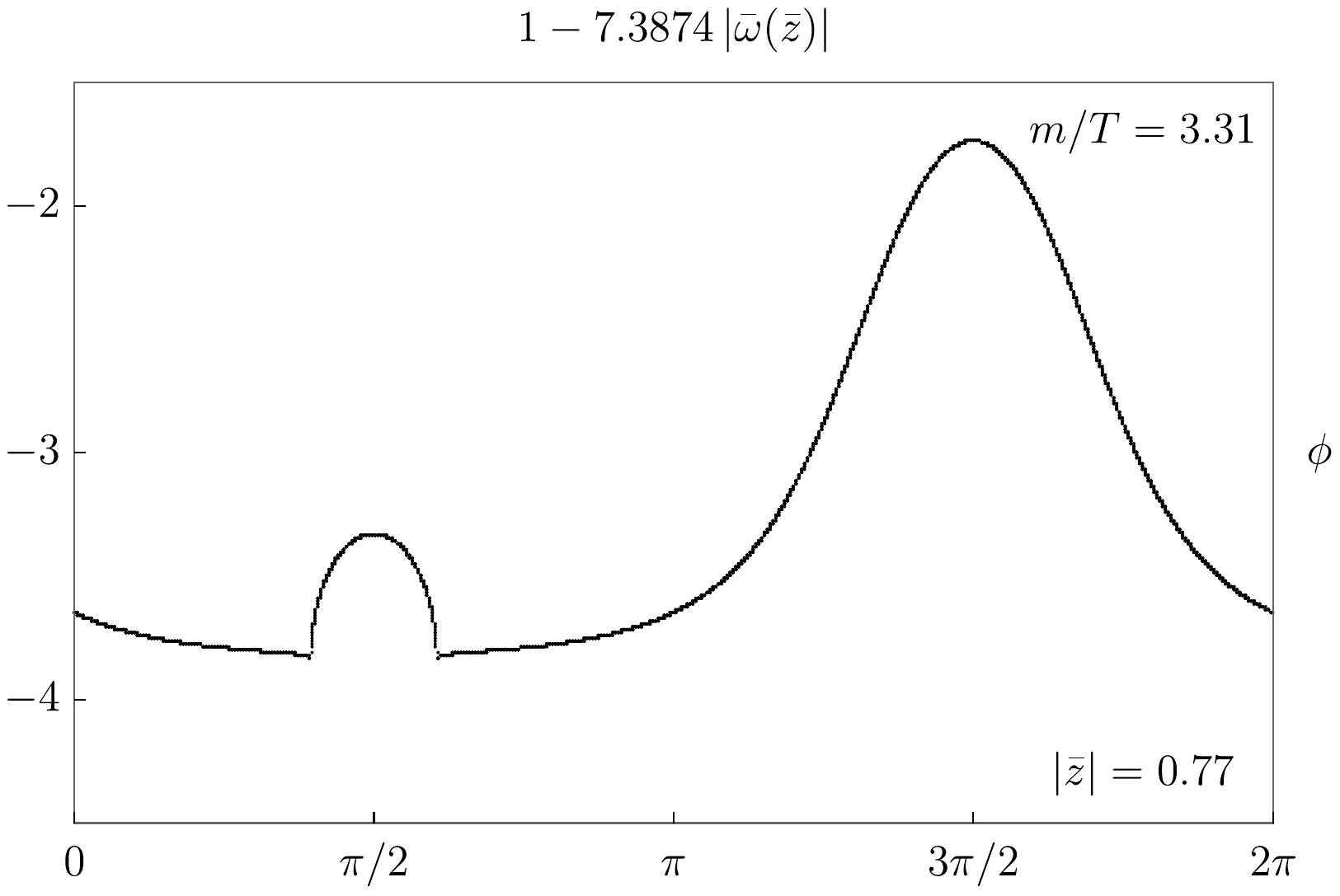}
    \caption{A check of the sufficient condition \eqref{con1} for the conformal bound on the speed of sound \eqref{ConformalBound} for the transverse sound mode in Case I${}_T$ (\textbf{left panel}) and the longitudinal sound mode in Case I${}_L$ (\textbf{right panel}). The black line indicates the value of the function \eqref{con2} with respect to the argument of $z$, i.e., $\phi \in [0, 2 \pi]$, specified for the relevant parameters above each plot. Every $\phi$ where the function crosses a zero (plotted with a red dashed line) is one $\phi_0$ in Eq.~\eqref{con2}. Therefore, we can see that the transverse mode depicted on the left panel satisfies the conformal bound whereas the longitudinal mode on the right may not (and does not).}
    \label{fig:conf}
\end{figure}

We test the expression \eqref{con2} on two examples: the transverse sound mode in Case I${}_T$ with $r = |z_g| = 2 \pi T |\bar z_g| $ that satisfies the conformal bound (cf.~Eq.~\eqref{TsoundSpeed}) and the longitudinal sound mode in Case I${}_L$ with $r = R_s = 2 \pi T \bar R_s $ that violates the conformal bound (cf.~Eq.~\eqref{LsoundSpeed}). In Figure~\ref{fig:conf}, we show that, indeed, the condition \eqref{con2} is satisfied by (two values of) $\phi_0$ for the transverse mode, while for the longitudinal mode, the condition fails to be satisfied.

Motivated by the fact that the speed of sound in Case I${}_L$ is outside the conformal bound regime, we can also devise a sufficient univalent condition by using the bounds in Eq.~\eqref{v1} to test whether a speed of sound is outside the conformal bound regime. In particular, for the bounds, we write
\begin{equation}\label{ConformalViolation}
v_c \leq v_s \leq \CC,
\end{equation}
where $\CC$ is some unspecified upper bound. It is easy to see that in order for \eqref{v1} to become \eqref{ConformalViolation}, we must look for a point along the circle $\zeta_0 = |\zeta_0| e^{i \phi}$ in the univalent $\zeta$ plane, such that 
\begin{equation}\label{Viol_zeta}
|\zeta_0| = \frac{\sqrt{\CC} - \left(\frac{1}{d-1} \right)^{1/4}}{\sqrt{\CC} + \left(\frac{1}{d-1} \right)^{1/4}},
\end{equation}
which satisfies the following condition:
\begin{equation}\label{con1-violation}
    |\partial_\zeta \varphi^{-1}(0)|= \frac{4 |\omega_0(z_0)|}{\CC - \sqrt{\frac{1}{d-1}}} =  \frac{4  \left|\omega_0 \! \left( \varphi^{-1}( |\zeta_0| e^{i\phi_0})\right)\right|}{\CC - \sqrt{\frac{1}{d-1}}} .
\end{equation}
In our examples, this translates into checking the existence of the following relation:
\begin{equation}\label{cond2_viol}
    1 -  \frac{8}{\left(2 \CC - \sqrt{2} \right) r}  \left|\omega_0 \! \left( \frac{2 \sqrt{\CC} - 2^{3/4}}{2  \sqrt{\CC}  + 2^{3/4}} r e^{i\phi_0} \right)\right| = 0.
\end{equation}

Firstly, as a very crude check of this relation, we can use the fact that if $\CC \approx 1$ (the speed of light), then according to Eq.~\eqref{Viol_zeta}, $|\zeta_0| \approx 0.086$, which is small compared to one (the edge of the univalence disk after rescaling from $\mathbb{D}_r$). Hence, one may expect the full dispersion relation to be reasonably well approximated by a truncated hydrodynamic series to some `small' order. In the most extreme case, if we approximate $\omega(z)$ by only the ideal hydrodynamic term that depends on the speed of sound, $\omega = v_s z$, then the relation is satisfied for the following speed: \begin{equation}
v_s = \frac{1}{8} \left( \sqrt{2} + 2^{7/4} \sqrt{\CC} + 2 \CC \right).
\end{equation}
For $\CC \approx 1$, this expression gives $v_s \approx 0.847$, which indeed lies outside the conformal bound range: $(v_c \approx 0.707) < (v_s \approx 0.847) < 1$.

\begin{figure}[h]
    \centering
    \includegraphics[width=0.45\linewidth]{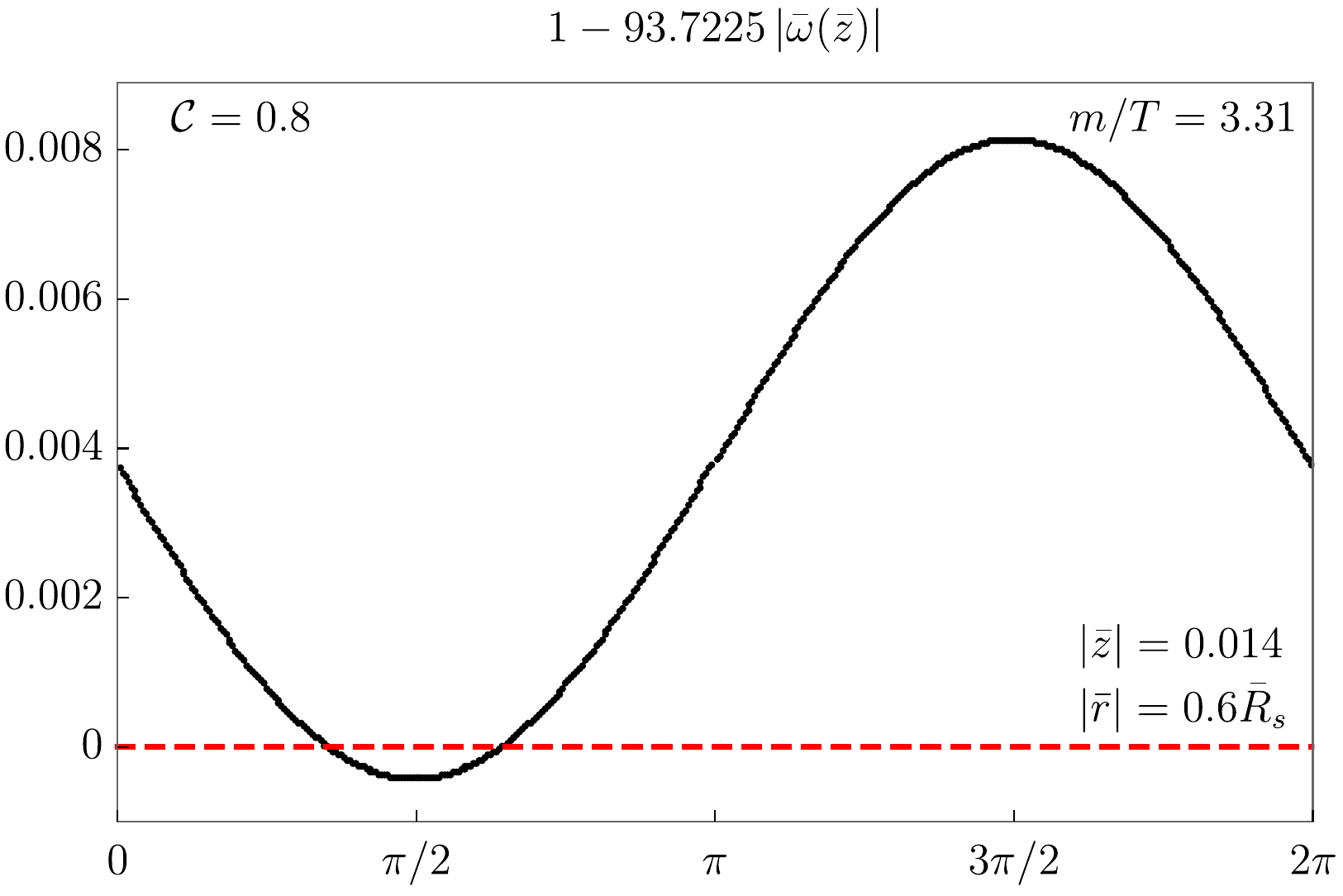}
    \caption{A check of the sufficient condition \eqref{con1-violation} for establishing the violation of the conformal bound on the speed of sound applied to the longitudinal sound mode in Case I${}_L$. The black line indicates the value of the function \eqref{cond2_viol} with respect to the argument of $\bar z$, i.e., $\phi \in [0, 2 \pi]$. The chosen values of parameters are $\CC = 0.8$ and $|\bar r| = |r| / (2\pi T) = 0.6 \bar R_s$ ($|\bar z| = 0.014$).}
    \label{fig:ConformalBoundViolation}
\end{figure}

Finally, we observe that the sufficient condition \eqref{cond2_viol} can indeed be satisfied in Case I${}_L$ for certain choices of $r < R_s$ and $\CC$. Therefore, the univalence condition can be correlated with the fact that $v_s$ lies outside of the conformal bound range. We present this for a specific choice of $r$ and $\CC$ in Figure~\ref{fig:ConformalBoundViolation}. We also note that the success of this particular condition is sensitively dependent on the choice of $\CC$, which cannot be too large. 

\section{Summary of observations}
\label{sec:Summary}

In this work, we tested a set of complex analytic methods and inequalities developed in Ref.~\cite{Grozdanov:2020koi} that can be used to derive rigorous and strict bounds on various hydrodynamic transport coefficients. These bounds are based on the theory of univalent (complex, holomorphic and injective) functions. Since the crucial new ingredient in these considerations is injectivity of a hydrodynamic dispersion relation, which is a property that is poorly understood and for which there is little obvious physical intuition, our aim was to start building this necessary intuition and begin noticing general behaviour of univalence in hydrodynamics. For this purpose, we performed a number of non-trivial tests in the family of holographic `axion' models with either explicitly or spontaneously broken translational invariance. We mainly focused on studying the sizes of univalent regions and checking the resulting inequalities. 

Here, we collect some of the main observations for diffusive and sound modes based on our examination of the holographic models in Sections~\ref{sec:Diff} and \ref{sec:solids}, as well as those from Ref.~\cite{Grozdanov:2020koi} for translationally invariant thermal CFTs. Despite their appealing simplicity and utility, we stress that for the moment, these statements remain unproven and should therefore not be considered as general.

In all studied examples of diffusion, we observed $\omega(z = q^2)$ to be univalent on the entire disk of hydrodynamic series convergence $\mathbb{D}_R$ by satisfying the property that $\re[i \omega'(z)] > 0$. Therefore, after the rescaling from the disk $|z| < R$ to the unit disk $\mathbb{D}$ with $|\zeta| < 1$, the condition $\re f'(\zeta) > 0$ (cf.~Eq.~\eqref{MacGregor}) is always satisfied. This implies that either the bounds \eqref{GrowthTheorem} and \eqref{deBranges} or the stronger {\it log} bounds \eqref{GrowthMacGregor} and \eqref{MacGregorCoeffs} can be used. Moreover, we also observed $\omega(z)$ to be univalent over `large' regions of $z\in\mathbb{C}$ (larger than $\mathbb{D}_R$), but not over the entire complex plane (or rather, the Riemann surface). In these cases, $\omega(z)$ are the exact dispersion relations, which should be considered as the analytic continuations of the hydrodynamic series \eqref{WDiff}. The special case of energy diffusion in the self-dual linear axion model studied in~\cite{Grozdanov:2020koi}, which is univalent on the entire complex plane with the exception of a semi-infinite cut, is an exception. Concrete examples of bounds on energy diffusion in the linear axion model and higher-order corrections to the dispersion relation were studied in Section~\ref{sec:Diff}. Examples of diffusion in the longitudinal channel of the holographic solid model were considered in Section~\ref{sec:Solid_Long}. In all studied cases, the univalence bounds were satisfied by explicitly calculated transport coefficients. Unsurprisingly, we observed that the bounds become stricter when a larger region $U$ is considered. As for the choice of the known point $\omega_0(z_0)$, it is presumably beneficial to choose (if possible) a point that is close to the origin $z=0$. 

In all studied examples of sound modes, we observe $\omega(z = q)$ to be univalent on the disk $|z| < r$ where $r$ is the smaller value of either the radius of convergence of the hydrodynamic series $R$ or the local univalence breaking point $|z_g|$ where the group velocity of the mode $v_g (z_g)$ vanishes. Univalence is ensured by the condition $\re[\omega'(z)] > 0$. As in the diffusive examples, this implies that either the bounds \eqref{GrowthTheorem} and \eqref{deBranges} or the stronger {\it log} bounds \eqref{GrowthMacGregor} and \eqref{MacGregorCoeffs} can be used. The momentum $z_g$ was always found to be purely imaginary. We discuss this in greater detail in Appendix~\ref{appendix:Cusps}. In the case of sound modes, we also considered specific examples of bounds and showed that they were satisfied in the studied longitudinal and transverse channels of the holographic solid model (see Sections~\ref{sec:Solid_Long} and \ref{sec:Solid_Trans}). Finally, in Section~\ref{sec:Conf}, we considered the conformal bound on the speed of sound \eqref{ConformalBound} \cite{Hohler:2009tv,Cherman:2009tw} and showed that a concrete (sufficient) univalence condition derived in Ref.~\cite{Grozdanov:2020koi} could be successfully used to relate the univalence properties of a dispersion relation to the conformal bound. Whether the same condition is also satisfied by all other examples of theories that satisfy the conformal bound remains an open problem. On the other hand, what we also saw was that a significant amount of `fine-tuning' was needed to develop a similar univalence condition that could demonstrate the violation of the conformal bound. We see this as an important reminder that such complications may in some cases be unavoidable when dealing with sufficient conditions. Nevertheless, with all the relevant caveats and open problems in mind, our observations still give us some hope that in the future, we may be able to use the univalence methods in order to understand the conformal bound, as well as other bounds on transport, from a more fundamental perspective through the univalence properties of QFTs.  

Finally, we stress that as far as the mathematical framework discussed here is concerned, the use of these methods is not restricted to the hydrodynamic effective field theory (EFT). In fact, one should be able to use them in any EFT so long as one can argue that a certain quantity is univalent. Moreover, if that particular quantity is expressed in terms of some series expansion, for example, a gradient expansion (as in hydrodynamics) or a Wilsonian EFT expansion, then the univalence methods can be used to derive bounds on the coefficients parametrising the series. Recently, this was for example done for a Wilsonian expansion of crossing-symmetric scattering amplitudes in Ref.~\cite{Haldar:2021rri}. We believe that such methods will indeed be broadly useful for constraining quantum and classical effective field theories in the future.

\acknowledgments{Matteo Baggioli acknowledges the support of the Shanghai Municipal Science and Technology Major Project (Grant No.2019SHZDZX01) and the sponsorship from the Yangyang Development Fund. M.B. would like to thank IFT and UAM Madrid for the warm hospitality during the completion of this work. Sebastian Grieninger is in part supported by the ‘Juan de la Cierva-Formación' FJC2020-044057-I program funded by MCIN/AEI/10.13039/ 501100011033 and NextGenerationEU/PRTR, and the `Atracci\'on de Talento' program (2017-T1/TIC-5258, Comunidad de Madrid) as well as through the grants CEX2020-001007-S and PGC2018-095976-B-C21, funded by MCIN/AEI/10.13039/501100011033 and by the ERDF A way of making Europe. The work of Sa\v{s}o Grozdanov is supported by the STFC Ernest Rutherford Fellowship ST/T00388X/1 and the research programme P1-0402 of Slovenian Research Agency (ARRS). Zhenkang Lu wants to thank Prof.~Shaofeng Wu for his supervision on related calculations in the linear axion model and Prof.~Andrzej Rostworowski for helpful discussions on constructing master fields.}

\begin{appendix}
\section{Local breaking of univalence in sound modes}
\label{appendix:Cusps}
    
In this appendix, we discuss in more detail the  phenomenon of local univalence breaking at $z_g$ (cf.~Eq.~\eqref{local_cond}). In terms of the complex spectral curve, the existence of such a point implies a formation of a cusp or a self-intersection along the dispersion relation curve $\omega(z)$ in the $z$ plane (see Figures~\ref{fig:CuspN4SYM} and \ref{selfself}). Here, we will focus only on sound modes where $z_g$ often plays an important role in limiting univalence. In particular, the existence of $z_g$ inside the radius of convergence limits the size of a univalent disk $U$ centred at $z=0$, which is the simplest $U$ that one can consider. 

\begin{figure}[h!]
\centering
 \includegraphics[width=0.65\textwidth]{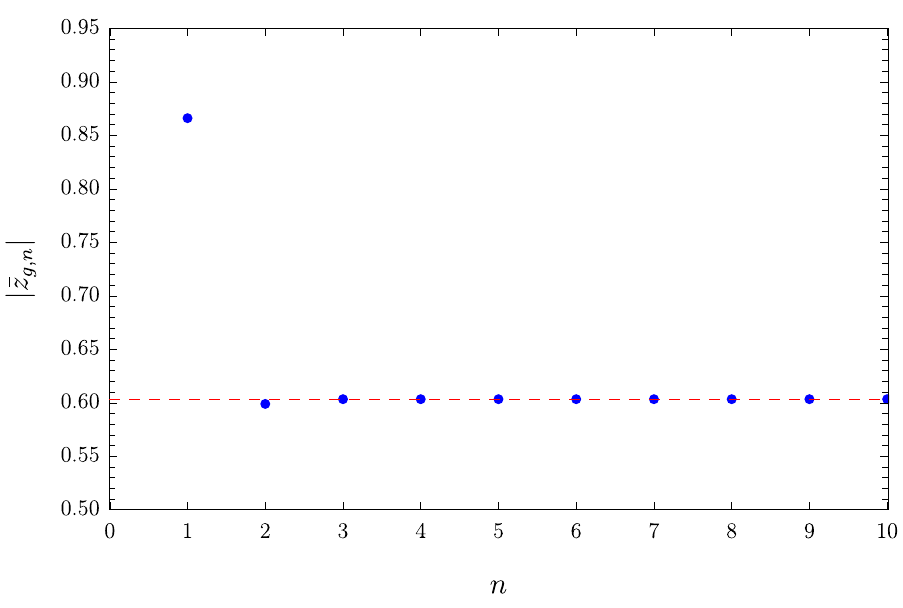}
\caption{A plot showing the rapid convergence of the Newton-Raphson expression in Eq.~\eqref{equ:NRiteration} for $|\bar z_{g}|$ in the $\CN =4 $ SYM theory plotted for orders $n = \{1, \ldots, 10\}$. The red dashed line indicates the exact (numerically computed) value of $|\bar z_g| = 0.603$ (see Figure~\ref{fig:CuspN4SYM}).} 
\label{fig:NRSYM4}
\end{figure}

We start by discussing the observation that in cases studied so far, $z_g = q_g$ is an imaginary (analytically continued absolute value of the) wavevector. This further implies that $\omega(z_g)$ is also imaginary. One interesting way to understand this fact is by first truncating the sound dispersion relation series \eqref{WSound} $\omega(z)$ (note that we take $\omega = \omega_+$) at order $z^2$ by defining 
\begin{equation}\label{omega_trunc}
\omega_{\rm trunc}(z) \equiv a_1 z + i a_2 z^2,
\end{equation} 
with $a_n \in \mathbb{R}$, $a_1 > 0$ and $a_2<0$. Equation $\omega_{\rm trunc}'(z) = 0$ that gives $z$ for which the group velocity of the truncated mode vanishes is now solved by 
\begin{equation}
z_{g,1} = \frac{i a_1}{ 2 a_2 },
\end{equation} 
which is always imaginary. We call this solution by the order of hydrodynamics at a given truncation --- in this case, first-order hydrodynamics. Starting from $z_{g,1}$, one can then by iteration construct a root of $\omega'(z) = 0$ by the Newton-Raphson method:
\begin{equation}
\label{equ:NRiteration}
 z_{g,n} = z_{g,n-1} - \frac{\omega'(z_{g,n-1})}{\omega''(z_{g,n-1})},
\end{equation}
where $n \geq 2$. Given the structure of the full series $\omega$ in Eq.~\eqref{WSound} and that $z_{g,1}$ is imaginary, it immediately follows that all $z_{g,n}$ are imaginary. We expect $\lim_{n\to\infty} z_{g,n}$ to (typically) converge to the exact value of $z_g$ if $z_g$ defined by the full dispersion relation $\omega(z)$ exists within the radius of convergence of the hydrodynamic series. It is easy to check that, for example, $z_{g,n\to\infty}$ in the $\CN = 4$ SYM theory indeed converges to the right value of $z_g = - 0.603 (2\pi T) i$ (cf.~Figure~\ref{fig:CuspN4SYM}). We show this in Figure~\ref{fig:NRSYM4}. 

\begin{figure}[h!]
 \includegraphics[width=1\textwidth]{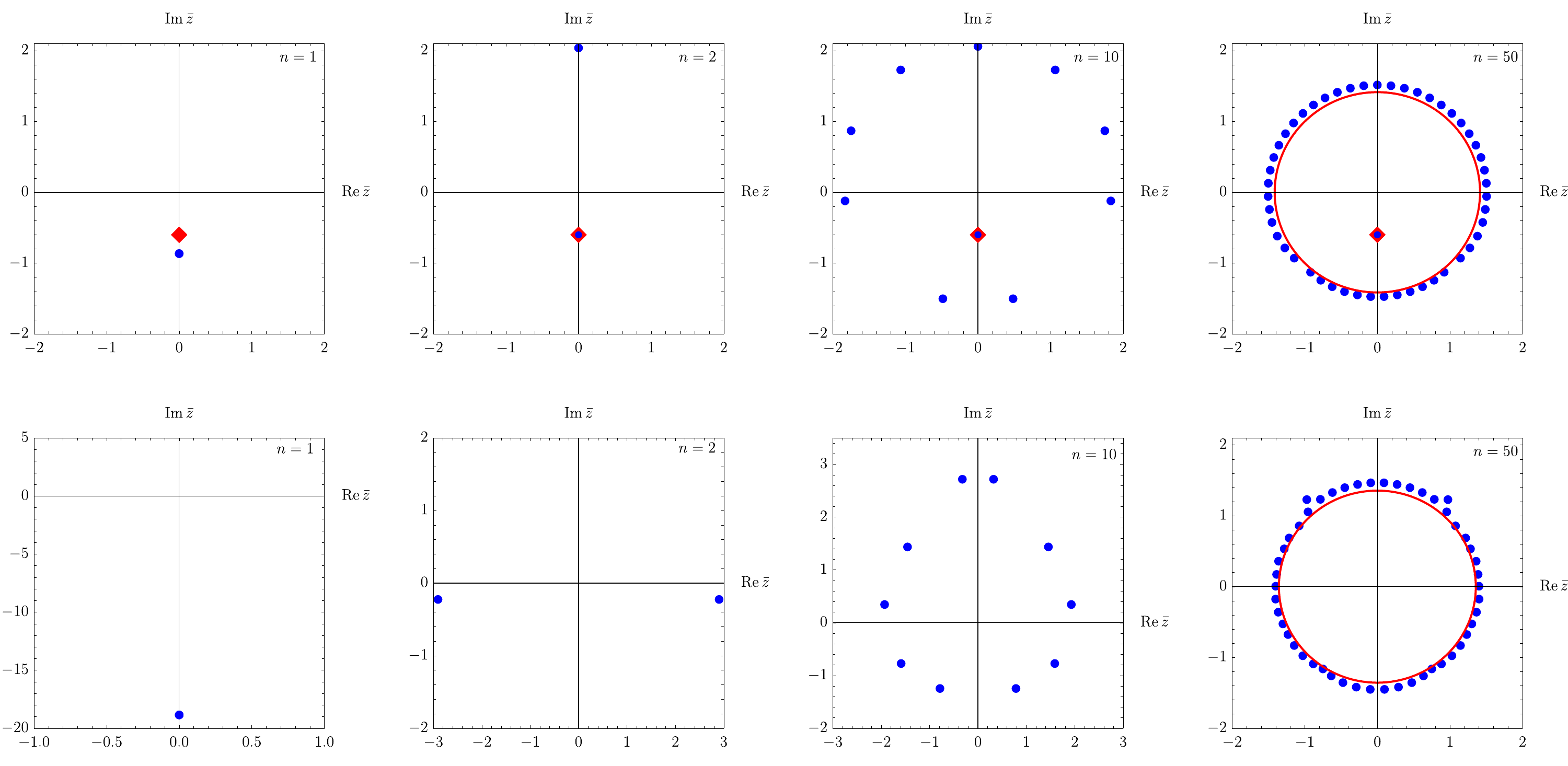}
\caption{The distribution of roots of the equation $\bar{\omega}'(\bar{z})=0$ truncated at orders $n = \{1,2,10,50\}$ of the hydrodynamic gradient expansion. \textbf{Top panel:} Here, we show the example of the longitudinal sound mode in $\CN = 4$ SYM theory. In the four plots, the red diamond represents the location of the self-intersection point $\bar{z}_g$. We see that already at order $n=2$, the location of one of the roots is virtually indistinguishable from the exact value of $\bar{z}_g$.  The spurious poles organise themselves along the radius of convergence $|\bar z| = \bar R = \sqrt{2}$, depicted as a red circle in the last plot ($n = 50$). \textbf{Bottom panel:} Here, we show the example of the Case III${}_L$ of the holographic solid model (cf.~Table~\ref{tab2}). The spurious roots (all roots) converge to the boundary of $\mathbb{D}_R$, with $\bar R = 1.36$, again depicted as a red circle in the last plot ($n=50$).} 
\label{fig:roots}
\end{figure}

In the spirit of the analysis done in Section 5 of Ref.~\cite{Grozdanov:2021gzh}, it is also interesting to study the behaviour of all (non-perturbative) roots of the polynomial equations given by $\omega'(z)$ at different orders of truncation. We again label the order of truncation by the order $n$ of the hydrodynamic expansion it corresponds to (ideal is zeroth-order, viscous is first-order giving \eqref{omega_trunc}, and so on). What we observe is that in the $\CN = 4$ SYM theory, with $|z_g| < R = 2 \sqrt{2} \pi T $ \cite{Grozdanov:2019kge}, one of the roots converges to the value of $z_g$. In fact, similarly to our observation in Figure~\ref{fig:NRSYM4}, this convergence is also very fast. For example, if we truncate the sound dispersion relation at order $z^3 = q^3$ (see e.g.~Refs.~\cite{Grozdanov:2015kqa,Grozdanov:2019uhi}), then we can compute $z_g$ analytically. We find $\bar z_g = -0.608 i$, which is in very good agreement with the actual value of $\bar z_g = - 0.603 i$. As $n$ grows, the other spurious (unphysical) roots organise themselves along the circle at the boundary of the convergence disk $\mathbb{D}_R$. On the other hand, in Case III${}_L$, where $|z_g| > R$, only the set of spurious roots organised at the edge of the disk of convergence remains. We show these phenomena for truncations at orders $n = \{1,2,10,50\}$ in Figure~\ref{fig:roots}.

\section{Collision between diffusive and sound hydrodynamic modes and the radius of convergence}\label{appendix:Coll}

\begin{figure}[ht]
    \centering \includegraphics[width=0.48\linewidth]{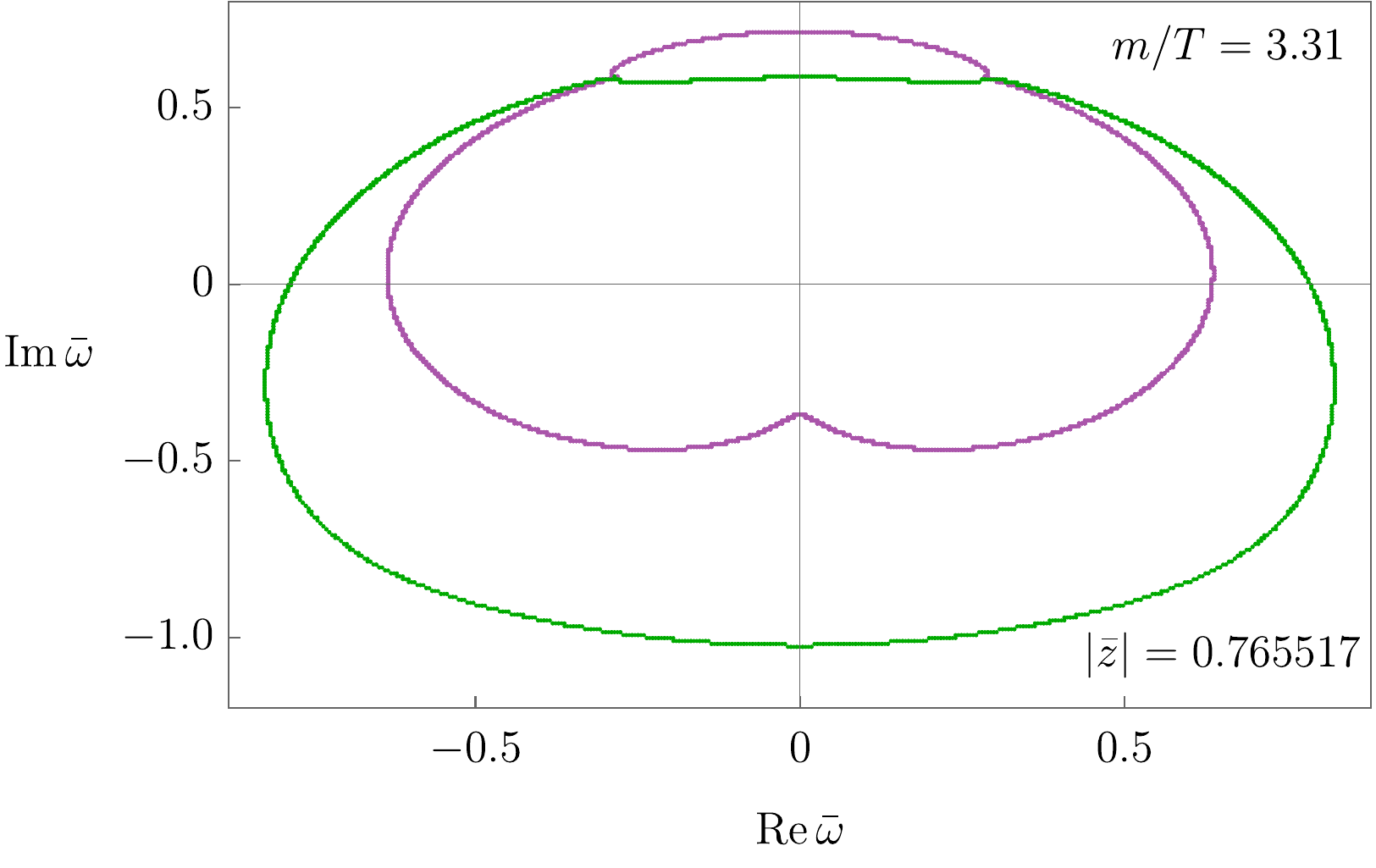}
    \includegraphics[width=0.48\linewidth]{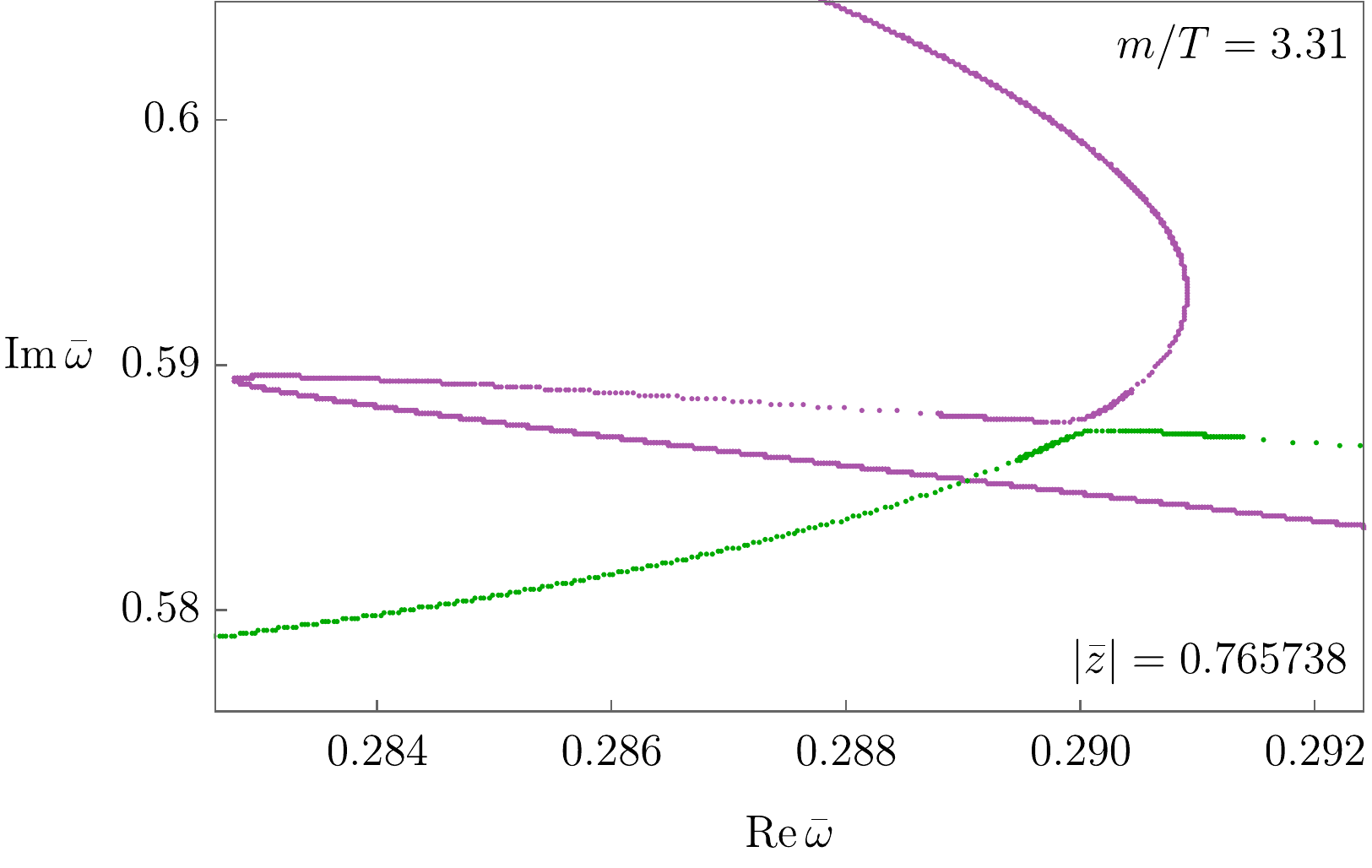}\newline
    \includegraphics[width=0.48\linewidth]{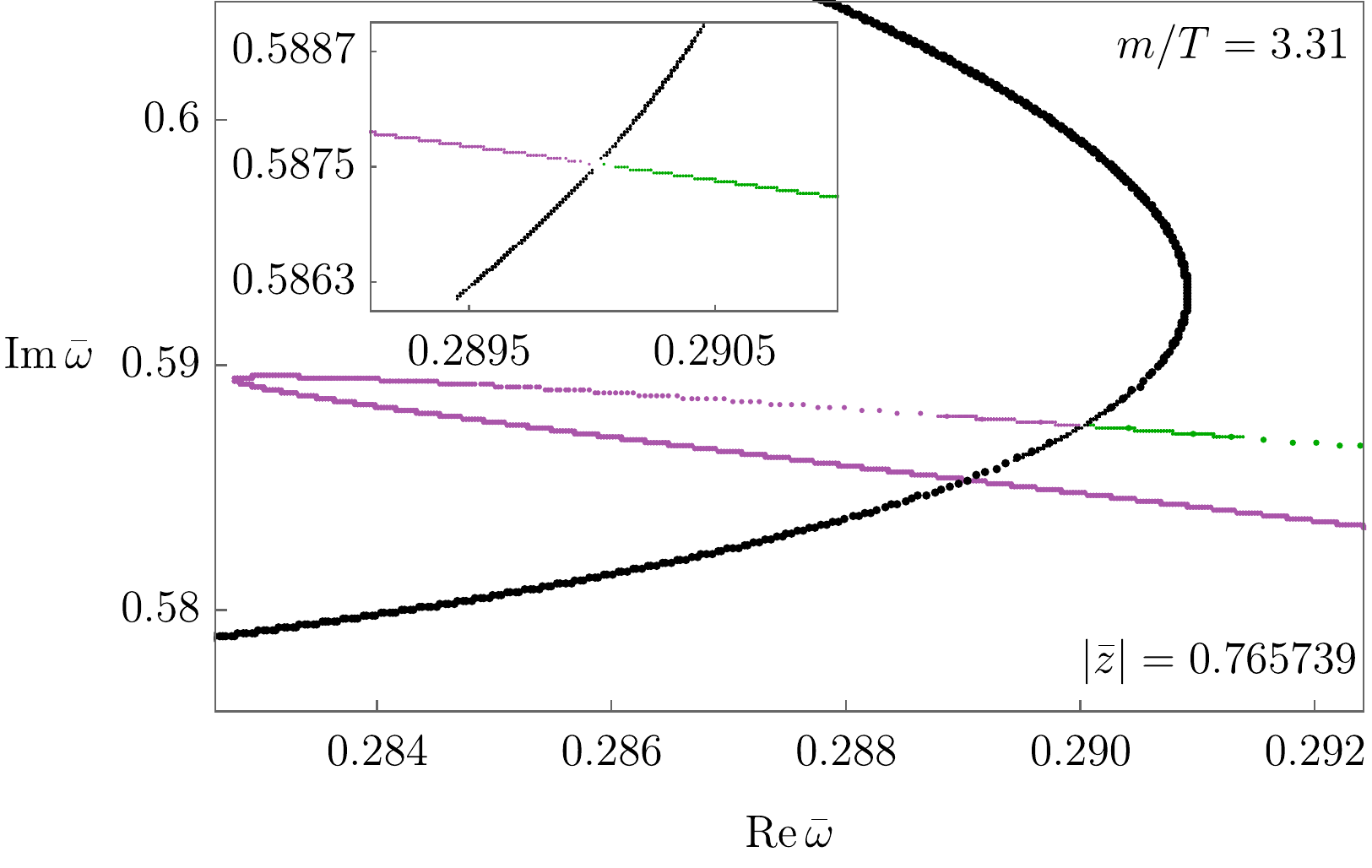}
    \includegraphics[width=0.48\linewidth]{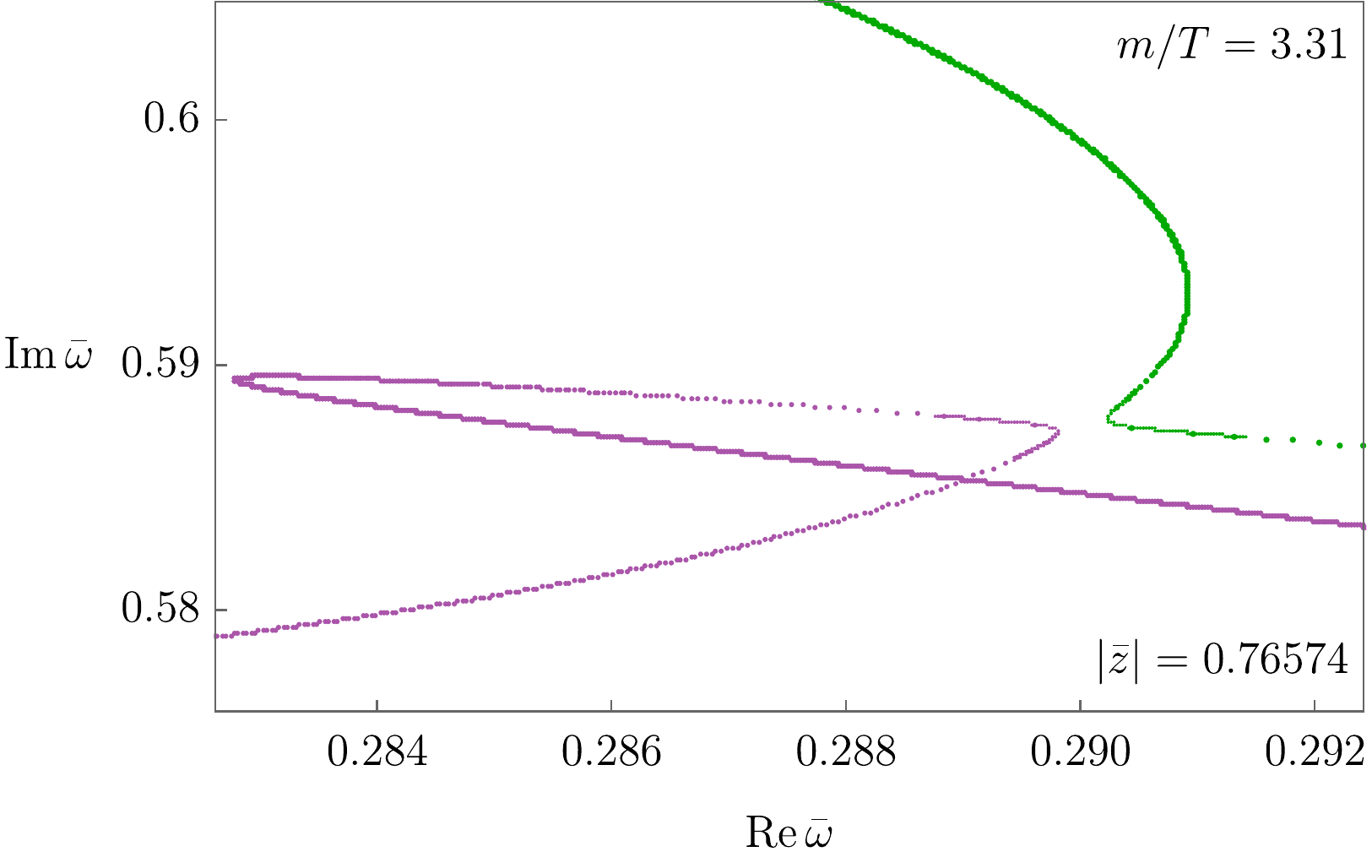}     
    \caption{Depiction of the collision between the hydrodynamic diffusive and sound modes in the longitudinal channel of the holographic solid model at $m/T=3.31$ (Case I${}_L$), plotting $\bar\omega(\bar z)$ as a function of $\phi$ in $\bar z = |\bar z| e^{i\phi}$. The green colour corresponds to the diffusive mode and the purple colour corresponds to the sound modes. The top two plots represent the modes' trajectories before the collision (a zoomed-out and a zoomed-in version at slightly different $|\bar z|$). At the value of $|\bar z|$ that results in a collision (a critical point), which is depicted in the bottom left plot, the curves are marked in black after they collide because the two hydrodynamic modes cannot be distinguished. The bottom right plot shows level-crossing of the two modes for $|\bar z|$ larger than that of the critical point.}
    \label{fig:colli}
\end{figure}

Finally, in this appendix, we provide additional details about the statements made in Section~\ref{sec:solids} that pertain to Case I${}_L$. In particular, we show that the radii of convergence of the coupled diffusive and sound modes are equal (cf.~Table~\ref{tab2}). This is because of the collision between the two types of hydrodynamic modes, which sets the lowest critical point of the spectral curve \cite{Grozdanov:2019kge,Grozdanov:2019uhi}. While collisions between diffusive and sound modes had been observed before, for example in the case of the charged Reissner-Nordst\"{o}m black branes \cite{Abbasi:2020ykq,Jansen:2020hfd}, in those cases, they give rise to subleading critical points that do not set the hydrodynamic radii of convergence. To the best of our knowledge, the longitudinal channel of the holographic solid model is the first example of such a collision. We present our numerical results that depict the collision for $m/T = 3.31$ in Figure~\ref{fig:colli}.

In this model, the phenomenon described here appears to be robust for small values of $m/T$. We observe that this type of collision persists all the way to the limit of $m/T \to 0$. In fact, in that case, the critical wavevector tends towards becoming purely imaginary with the value $\bar z = i/(2\pi T)$. Instead, by increasing the value of $m/T$, the non-hydrodynamic gapped modes move closer towards the origin of the complex frequency plane. In the limit of small temperatures ($m/T \gg 1$), the crystal diffusive mode collides first with the lowest of the purely imaginary non-hydrodynamic modes, which are a signature of the emergent AdS$_2 \times \mathbb{R}^2$ geometry (see e.g.~Refs.~\cite{Arean:2020eus,Wu:2021mkk}).

\section{Radius of convergence in the transverse sector of a holographic solid model}\label{appendix:R}

\begin{figure}[ht]
    \centering \includegraphics[width=0.6\linewidth]{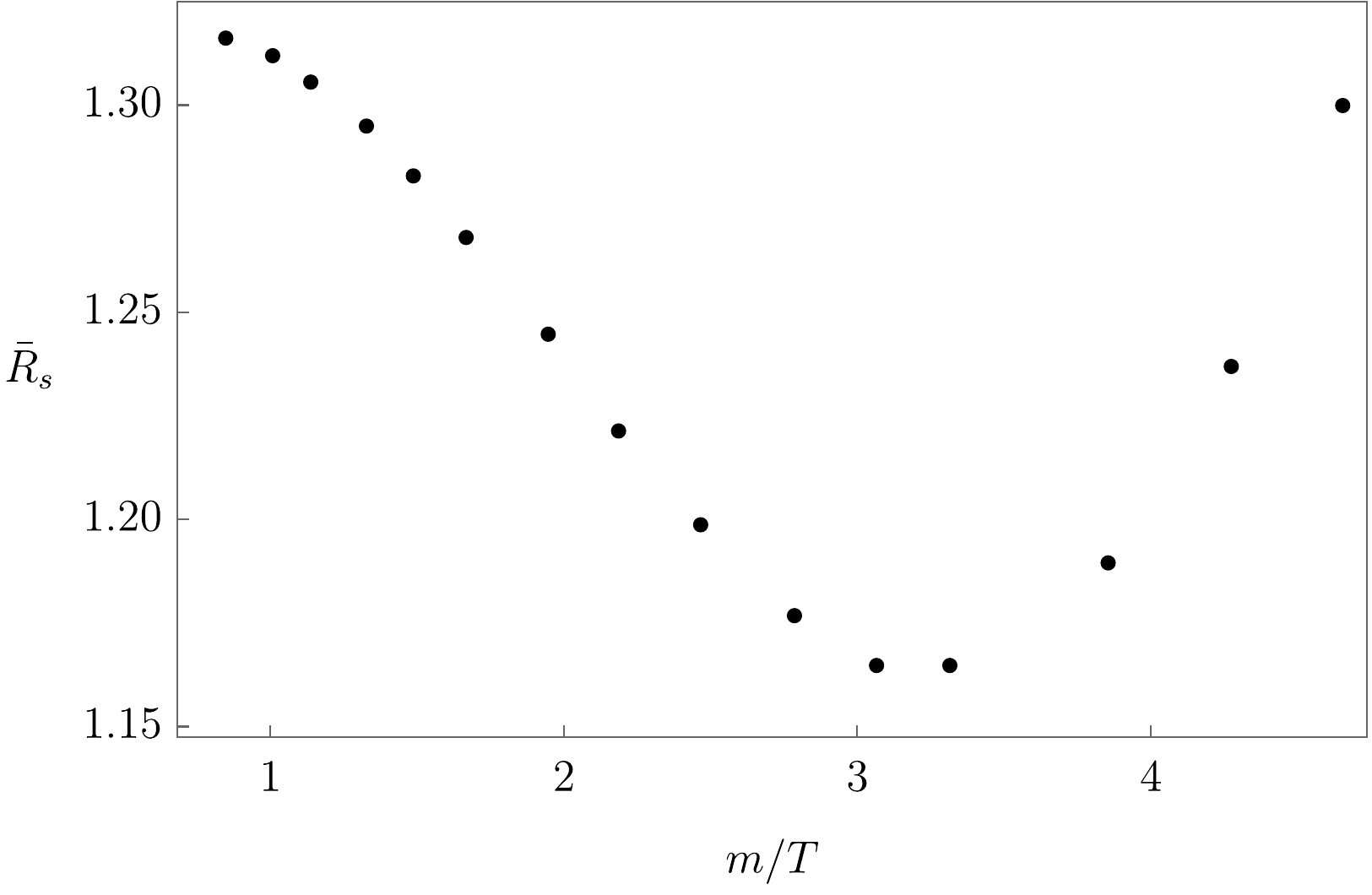}     
    \caption{Dimensionless radius of convergence $\bar{R}_s$ plotted as a function of the symmetry breaking parameter $m/T$ for holographic solids (with spontaneously broken translations) with the potential $V(X)=X^3$.}
    \label{fig:radcon}
\end{figure}

In this appendix, we show the radius of convergence $R_s$ plotted as a function of the strength of spontaneous symmetry breaking in the transverse sector (i.e., in the holographic solid model). As is evident from Figure~\ref{fig:radcon}, the radius of convergence does not change monotonically with the symmetry breaking scale. On the contrary, it first decreases for small values of $m/T$, reaches a minimum value (which depends on the chosen potential $V(X)$) and then grows at larger values of $m/T$. It would be interesting to understand this behaviour of $R_s$ and its physical meaning better and to extend this computation to the longitudinal sector, as well as to a larger class of potentials. We defer these problems to future research.

\end{appendix}

\bibliographystyle{JHEP}
\bibliography{Genbib.bib}
\end{document}